\documentstyle[12pt,epsfig]{article}
\def\beq{\begin{equation}}   \def\eeq{\end{equation}}
\def\bea{\begin{eqnarray}}   \def\eea{\end{eqnarray}}

\newcommand{\gsim}{\lower.7ex\hbox{$
\;\stackrel{\textstyle>}{\sim}\;$}}
\newcommand{\lsim}{\lower.7ex\hbox{$
\;\stackrel{\textstyle<}{\sim}\;$}}

\newcommand{\mvec}[1]{|\vec{#1}\,|}

\newcommand{\La}{\overline{\Lambda}}

\newcommand{\bibit}[1]{\bibitem{#1}}

\newcommand{\GeV}{\,\mbox{GeV}}
\newcommand{\MeV}{\,\mbox{MeV}}

\newcommand{\matel}[3]{\left< #1\left|#2\right|#3\right>}
\newcommand{\aver}[1]{\left< #1 \right>}

\newcommand{\state}[1]{ |#1\rangle}

\renewcommand{\sp}{ {\rm sp} }

\begin{document}

\def\lsim{\mathrel{\rlap{\lower3pt\hbox{\hskip0pt$\sim$}}
    \raise1pt\hbox{$<$}}}         
\def\gsim{\mathrel{\rlap{\lower4pt\hbox{\hskip1pt$\sim$}}
    \raise1pt\hbox{$>$}}}         

\count255=\time\divide\count255 by 60 \xdef\hourmin{\number\count255}
  \multiply\count255 by-60\advance\count255 by\time
  \xdef\hourmin{\hourmin:\ifnum\count255<10 0\fi\the\count255}

\begin{titlepage}
\renewcommand{\thefootnote}{\fnsymbol{footnote}}

\begin{flushright}
JLAB-THY-00-21\\
UND-HEP-00-BIG\hspace*{.1em}05
\end{flushright}
\vspace{.3cm}

\begin{center} \Large
{\bf Precision Studies of Duality in the 't~Hooft Model}
\end{center}
\vspace*{.3cm}
\begin{center}
{\Large
Richard F. Lebed$^{\:a}$\footnote{lebed@jlab.org} and Nikolai
G. Uraltsev$^{\:b,c}\footnote{uraltsev@undhep.hep.nd.edu}$
\\
\vspace{.4cm}
{\normalsize
$^a${\it Jefferson Lab, 12000 Jefferson Avenue, Newport News, VA
23606, U.S.A.}\\
$^b${\it Dept.\ of Physics,
Univ.\ of Notre Dame du
Lac, Notre Dame, IN 46556, U.S.A.}\\
$^c${\it Petersburg Nuclear Physics Institute,
Gatchina, St.\,Petersburg, 188350, Russia
}
}
}

\vskip 1cm
(June, 2000)
\vskip 1cm

\end{center}

\thispagestyle{empty}
\setcounter{page}{0}

\begin{abstract}
We address numerical aspects of local quark-hadron duality using the
example of the exactly solvable 't~Hooft model, two-dimensional QCD
with $N_c\to\infty$.  The primary focus of these studies is total
semileptonic decay widths relevant for extracting $|V_{cb}|$ and
$|V_{ub}|$.  We compare the exact channel-by-channel sum of exclusive
modes to the corresponding rates obtained in the standard $1/m_Q$
expansion arising from the Operator Product Expansion.  An impressive
agreement sets in unexpectedly early, immediately after the threshold
for the first hadronic excitation in the final state.  Yet even at
higher energy release it is possible to discern the seeds of
duality-violating oscillations.  We find the ``Small Velocity'' sum
rules to be exceptionally well saturated already by the first excited
state.  We also obtain a convincing degree of duality in the
differential distributions and in an analogue of $R_{e^+e^-}(s)$.
Finally, we discuss possible lessons for semileptonic decays of actual
heavy quarks in QCD.

\vspace*{.4cm}

\noindent
PACS numbers: 12.38.Aw, 11.10.Kk, 13.20.-v

\end{abstract}


\vspace*{.2cm}
\vfill
\end{titlepage}

\clearpage
\tableofcontents
\setcounter{page}{1}

\section{Introduction}

	Questions of how to compare hadronic observables to the
apparent underlying fundamental theory of QCD lie at the heart of
understanding the nature of strong interactions.  Thirty years after
its inception, QCD in $D\!=\!4$ spacetime dimensions still stubbornly
refuses to admit a global solution.  The asymptotic freedom property
of the theory permits the perturbative calculation of (Euclidean)
Green functions involving large values of momentum transfer or energy
release in terms of quarks and gluons, the fundamental objects of QCD.
But at lower scales one enters the nonperturbative regime, which not
only invalidates (or at least complicates) the standard perturbative
methods of field theory developed in QED, but also leads to a dramatic
change in the physical spectrum of the theory.  Instead of quarks and
gluons, only colorless hadrons are produced as asymptotic states in
any process, even at arbitrarily large energy.

	Many nontrivial theoretical techniques respecting QCD first
principles have been developed to study nonperturbative features of
the theory.  Yet despite numerous advances, no one has been able to
compute the masses, wavefunctions, or transition amplitudes of hadrons
in terms of quark masses and couplings directly from the QCD
Lagrangian.  Moreover, many existing theoretical tools are expressed
through various expansions in certain small parameters; the actual
range of each parameter where the expansions are applicable is often
not well known.  In such a situation, it is clearly advantageous to
build a soluble toy field theory that incorporates as many features of
the QCD Lagrangian as possible.

	Such a theory does indeed exist, the famous 't~Hooft model
\cite{thooft}, which is defined by the Yang-Mills Lagrangian in
$D\!=\!2$ spacetime dimensions in the limit of a large number of
colors $N_c$.  As was shown in the original paper, the quark-antiquark
sector of the theory admits an infinite tower of confined,
color-singlet solutions that can be obtained, in principle, to an
arbitrary degree of numerical accuracy.  The reason for this
solubility lies precisely in the defining features of the model.
Large $N_c$ eliminates all Feynman diagrams with internal $q\bar q$
loops and nonplanar gluons.  On the other hand, $D\!=\!2$ allows gluon
self-couplings to be eliminated by gauging away one component of the
gauge potential $A^\mu$.  Since only two components are initially
present, the commutator term $[ A^\mu , A^\nu]$ in the covariant
derivative, which gives gluon self-coupling, vanishes identically in
such gauges.  Then the only remaining Feynman diagrams to be summed
for the quark-antiquark Green function are ``rainbow'' and ``ladder''
diagrams, whose Schwinger-Dyson equations can be solved, giving rise
to an integral expression called the 't~Hooft equation (discussed in
Sec.~\ref{tHm}).

	The 't~Hooft model provides an excellent laboratory for
testing various approaches to strong interaction physics.  After all,
the 't~Hooft equation provides a means to compute hadronic masses,
wavefunctions, and transition amplitudes in terms of the underlying
partonic degrees of freedom.

	In this work we are specifically interested in questions of
local quark-hadron duality in the inclusive decays of heavy quarks.
The notion of duality in general terms was first introduced in the
early days of QCD in Ref.~\cite{poggio} but not pursued for quite some
time.  A more detailed consideration was given a few years ago by
Shifman \cite{shifdual} and later reiterated in a number of papers
(see, e.g., Refs.~\cite{inst,D2}), with applications relevant to
Minkowskian observables amenable to study via an operator product
expansion (OPE).  This allows the formulation of the concept of local
duality in a more quantitative way, including nontrivial
nonperturbative effects; we refer the reader to these recent
publications for the theoretical aspects.  Here the question of
duality is studied concretely by comparing the weak decay width of a
meson containing a heavy quark computed in two ways.  In terms of
partonic degrees of freedom, one has an OPE depending upon the free
quark diagram (with perturbative corrections) and a number of
nonperturbative matrix elements suppressed by powers of the heavy
quark mass.  In terms of the hadronic degrees of freedom, one simply
computes the weak decay amplitude for each allowed exclusive channel,
and adds them up one by one.  This comparison is especially
instructive since one may consider the behavior of solution as the
mass $m_Q$ of the heavy decaying quark is varied.

	Such a problem was first considered in Ref.~\cite{GL1}, where
the main elements in numerical computations of exclusive decay rates
were annunciated.  The hadronic result was compared to the Born-level
free-partonic diagram as a function of $m_Q$.  In terms of the OPE,
the latter is the tree-level piece of the Wilson coefficient
corresponding to the unit operator.  The numerical agreement was seen
to be remarkable, in that the onset of the asymptotic agreement was
clearly visible already for relatively small values of $m_Q$.  The
intrinsically limited numerical accuracy for sufficiently heavy
quarks, however, prohibited drawing a definite conclusion about the
size of nonperturbative corrections for asymptotically large $m_Q$.
Additional numerical studies \cite{GL2} considered similar questions
for weak decay topologies other than the simple spectator tree
diagram, in particular weak annihilation (WA).

	The validity of the OPE was addressed analytically in
Refs.~\cite{D2,D2WA,D2PI}, which considered on one hand the nature of
the OPE for heavy quark decays, and on the other an explicit $1/m_Q$
expansion of the decay amplitudes, which allows an analytical
summation of the individual decay rates in the asymptotic regime.  The
agreement of the two approaches through relative order $1/m_Q^4$ was
obtained by means of a number of sum rules derived directly from the
't~Hooft equation, the archetype of which first appeared in
Ref.~\cite{burk}.

	While adequate to illustrate the theoretical validity of the
OPE for the inclusive decay widths of heavy flavors, the analytic
methods per se cannot help in answering the practical question
relevant to phenomenology of beauty and charm quarks: Namely, how
accurately do the OPE-improved parton computations describe the true
weak decay width of a heavy flavor meson with finite mass, only a few
times larger than the typical strong interaction scale?  A purely
analytic expansion can hardly be used for this purpose, since it is a
priori unknown how small an expansion parameter must be for the
expansion to start yielding a reasonable approximation, not to mention
achieving the necessary precision.  To obtain insights into the size
of deviations between the actual decay widths and the expressions
obtained from the OPE for quarks in the intermediate mass range, one
must employ real numerical computations.

	In this paper we focus on semileptonic decays of heavy quarks.
In the contexts of both real QCD and the 't~Hooft model, they are
technically simpler than nonleptonic decays.  Moreover, the magnitude
of local duality violation is phenomenologically most important in
semileptonic decays when one extracts $|V_{cb}|$ and $|V_{ub}|$.  We
use the techniques developed in Ref.~\cite{GL1} to evaluate the
required decay rates, and confront the total decay width with the
expansion in terms of a power series in $1/m_Q$ of Ref.~\cite{D2}.
Moreover, by making use of a number of relations derived in the
large-$m_Q$ limit of the 't~Hooft equation \cite{burkur}, members of
the set of nonperturbative matrix elements involved can be related to
each other, providing an economical description of the nonperturbative
physics.  These are the tools that allow us to study the onset of
quark-hadron duality.

	As explained in Appendix~\ref{newmult}, we use a scheme based
on the modified Multhopp method, by which the 't~Hooft equation is
converted into an infinite-dimension eigenvector system that for
practical reasons must be truncated at some number $N$ of eigenvector
modes.  The asymptotic convergence of this approach has not been
rigorously studied, although it apparently must yield unlimited
accuracy when the number of the Multhopp modes $N$ goes to infinity.
Yet the rate of convergence at large $N$ is not well known.
Additionally, large quark masses turn out to require one to use a
larger $N$ for sufficient numerical accuracy, as discussed in
Sec.~\ref{tHm}.  It therefore seems mandatory to make an independent
cross-check of the numerical accuracy.  We investigate this problem by
comparing the numerical values of a number of static properties of
heavy mesons at different values of $m_Q$, with the results of their
$1/m_Q$ expansions obtained analytically from the 't~Hooft equation;
this is the topic of Sec.~\ref{check}.  We find that our solutions
have sufficient numerical accuracy for masses $m_Q$ corresponding to
physical values (in the sense explained in Sec.~\ref{check}) as large
as $20$\,GeV.

	The duality of the inclusive widths of heavy-flavor hadrons to
the parton-level widths, including the power corrections from the OPE,
emerges through a set of sum rules that equate sums of weighted
transition probabilities to possible final states and expectation
values of the local heavy quark operators.  Since our main interest
lies in $b\to c$ transitions, which carry in practice a limited energy
release, the most relevant are the so-called small velocity (SV) sum
rules, which we study here in the heavy quark limit.  The behavior of
these sum rules not only shapes the semileptonic $b\to c$ decays in
actual QCD, but is also important for the determination of the basic
parameters of the heavy quark expansion.

	An additional advantage of the heavy quark limit for our
investigation is that we are able to compute the SV amplitudes
semi-analytically, using the exact relations \cite{burkur} derived
from the 't~Hooft equations and relying for input only on a few static
parameters, which can be computed with a high precision.  A discussion
of these relations appears in Sec.~\ref{dual}.  We find that the SV
sum rules in the 't~Hooft model are saturated to an unexpectedly high
degree by the first excitation above the ground state (which we
henceforth call the ``$P$-wave'' excitation, despite the fact that in
$D\!=\!2$ only radial excitations occur).  Its contributions to even
the Darwin ($\rho_D^3$) and kinetic ($\mu_\pi^2$) expectation values
constitute over $90\%$ and $96\%$ of the totals, respectively, while
it saturates the ``optical'' sum rule for $M_B\!-\!m_b$ to a $1.5\%$
accuracy.  This appears to be an intriguing dynamical feature of the
model.  A similar high-saturation effect has been observed in a quark
flux-tube model \cite{Nathan}, for the contribution from the
``valence'' quarkonium states.

	We study the size of violations of local duality in the
semileptonic decays $b \to c \,\ell \bar\nu$ assuming vectorlike weak
currents and massless leptons.  These assumptions are important for
comparison with QCD far beyond the obvious parallel of closely
resembling the actual world: The strength of the resonance-related
duality violation crucially depends on the threshold behavior in the
decay probabilities, which is completely different in two and four
dimensions.  The two-body phase space, while $\propto \mvec{p}$ in
$D\!=\!4$, is $\propto 1/\mvec{p}$, that is, infinite at threshold, in
$D\!=\!2$.  On the other hand, the situation is special for massless
leptons: Their invariant mass is always zero if they are produced by a
vectorlike source, and the weak vertex is then proportional to the
momentum.  As a result, in this case the threshold behavior of the
decay rate becomes $\propto \mvec{p}$ much in the same way as in real
QCD.  This is a crucial detail if one tries to draw practical lessons
from the 't~Hooft model.  The need for a vectorlike coupling in
$D\!=\!2$ is even more stark for the parton-level calculation.  There
one finds that the integrated three-body phase space actually diverges
for massless leptons, and only the behavior of the weak decay
amplitude renders the width finite.  We provide more arguments in
favor of such a choice in Sec.~\ref{wid}, which is dedicated to the
inclusive decay widths.

	In Sec.~\ref{vac} we briefly illustrate how well the duality
works for the vacuum correlator of light quarks in the timelike
domain.  In the context of the heavy quark expansion this is relevant
for the nonleptonic decay widths, including spectator-dependent
effects like WA.

	Section~\ref{concl} summarizes our investigation and discusses
the conclusions that can be drawn for actual QCD.

	Appendices describe the computational technique employed and
contain a number of relations for the heavy quark limit of the
't~Hooft equation employed in these numerical studies.

\section{The 't~Hooft Equation and Its Solutions}
\label{tHm}

	We first review some well-known properties of the 't~Hooft
model both as a reminder and to establish notation.  Confinement is
manifest in 1+1 spacetime dimensions with large $N_c$, and the
quark($m_1$)-antiquark($\overline{m}_2$) two-particle irreducible
Green function, i.e., the meson wavefunction $\varphi(x)$, is given by
the 't~Hooft equation:
\begin{equation} \label{tHe}
M_n^2 \, \varphi_n (x) = \left( \frac{m_1^2 \!-\!
\beta^2}{x} + \frac{m_2^2 \!-\! \beta^2}{1\!-\!x} \right)
\varphi_n (x) - \beta^2 \int^1_0 {\rm d}y \:
\varphi_n (y) \, {\rm P} \frac{1}{(y\!-\!x)^2} \;,
\end{equation}
where $x$ is the momentum fraction in light-cone coordinates carried
by the quark, and
\begin{equation}
\beta^2 \equiv \frac{g_s^2}{2\pi} (N_c \!-\! 1/N_c) \; .
\end{equation}
Since $\beta$ is finite in the large-$N_c$ limit, it provides a
natural unit of mass.  Thus, all masses in this paper are understood
as multiples of $\beta$.  Indeed, as pointed out in Ref.~\cite{GL1},
$\beta$ fills the role in 1+1 dimensions of served by $\Lambda_{\rm
QCD}$ in 3+1.  We discuss the estimation of $\beta$ as a particular
number in Sec.~\ref{check}.

	The singularity of the QCD Coulomb interaction in
Eq.~(\ref{tHe}) is regularized using a principal value prescription,
indicated by P in Eq.~(\ref{tHe}).

	Solutions $n = 0, 1, \ldots$ of the 't~Hooft equation
alternate in parity, with the lowest being a pseudoscalar.  The
general analytic solution in closed form is not known.  As the
eigenvalue index $n$ increases, the eigenvalues $M_n^2$ asymptotically
approach $\beta^2 [\pi^2 n + O(\ln n)]$.

	The static limit $m_1 \equiv m_Q \rightarrow \infty$ is most
easily studied \cite{D2WA,burk,burkur} by employing the
``nonrelativistic'' variables $M_n\!=\!m_Q+\epsilon_n$,
$\,t\!=\!(1\!-\!x)m_Q$ and $\Psi_n(t)\!=\!\frac{1}{\sqrt{m_Q}}
\varphi_n\!\left(1\!-\!\frac{t}{m_Q}\right)$, in terms of which
Eq.~(\ref{tHe}) assumes the form
\begin{equation}
\epsilon_n \Psi_n(t) \;=\;
\frac{m_2^2\!-\!\beta^2}{2t} \Psi_n(t) + \frac{t}{2}
\Psi_n(t) -
\frac{\beta^2}{2}\, \int_{0}^{\infty}
{\rm d}s\:\frac{\Psi_n(s)}{(t\!-\!s)^2}\;\;.
\label{41n}
\end{equation}

        We solve the finite-mass 't~Hooft equation using a numerical
method called the Multhopp technique \cite{Mul}, a venerable system
for solving integral equations with singular kernels.  It was first
applied to the 't~Hooft equation in Ref.~\cite{HPP}.  The idea is to
expand the wavefunction in a series of modes, not unlike Fourier
analysis, and then turn the equations for the mode coefficients into
an equivalent infinite-dimension eigenvector problem.  In practice,
one then truncates at some point where the higher modes are deemed to
have little effect upon the wavefunction solutions, which is of course
strongly dependent on the highest value of $n$ used.  The detailed
formulas for applying the {\em standard\/} Multhopp technique to
mesons with unequal quark masses in the 't~Hooft model appear in
Appendix~A of Ref.~\cite{GL1}.

        Intrinsic to the original Multhopp technique is the evaluation
of the wavefunction at a discrete set of points called ``Multhopp
angles,'' which in the current problem are equivalent to
\begin{equation}
x_k = \frac 1 2 \left[ 1 + \cos {\left( \frac{k\pi}{N+1} \right)}
\right] \;,
\hspace{1em} \;\; k = 1, \ldots, N \; ,
\end{equation}
where $N$ is the number of modes retained in the numerical solution.
The mode coefficients are then obtained by the use of a discrete
inversion formula [(A7) in \cite{GL1}].  However, the Multhopp
solutions can be seen to vanish as $\sqrt{x}$ and $\sqrt{1\!-\!x}$ at
the endpoints $x\!=\!0$ and $x\!=\!1$, respectively [see (A10)--(A11)
in \cite{GL1}], while the exact solutions are known to vanish as
$x^{\gamma_1}$ and $(1\!-\!x)^{\gamma_2}$, respectively, where
\begin{equation}
\frac{m_i^2}{\beta^2} \,+\, \pi \gamma_i \cot {\pi \gamma_i}
\:=\: 1 \; ,
\end{equation}
leading to a type of Gibbs phenomenon in the Multhopp solutions.
Since the Mul\-thopp angles cease to sample the wavefunction at some
finite distance from the endpoints, it may be expected that the
wavefunctions thus obtained are numerically inaccurate there.  This
shortcoming led Brower, Spence, and Weis \cite{BSW} to improve the
Multhopp technique by eliminating the Multhopp angles and using
instead a continuous inversion formula.  The algebraic details are
presented in Appendix~\ref{newmult}, and it is this improved numerical
technique that is used in obtaining our results.

\section{Heavy Quark Expansion and Cross Check of the Algorithm}
\label{check}

	Let us first establish a bit of notation.  The mass of a heavy
quark of flavor $Q$ is labeled as $m_1 \to m_Q$; in the weak
transitions considered in subsequent sections, the final-state quark
$q$ is assigned the mass $m_q$.  The spectator antiquark mass $m_2$ is
labeled by $m$, or $m_{\rm sp}$ if there is any chance of confusion.

	As explained in the previous section and
Appendix~\ref{newmult}, we use the modified Mul\-thopp technique to
find numerical solutions of the 't~Hooft eigenstate problem.  Since
the heavy meson wavefunctions are peaked near the end of the interval,
the accuracy deteriorates with increasing $m_Q$.  The same, in
principle, applies to the high excitations of light hadrons.  A more
appropriate strategy for heavy quarks is to start with a solution of
the infinite-mass (static) equation.  This has been done analytically
\cite{D2,D2WA}, and full consistency\footnote{In the case that the
fermions $f$ created by the weak current have $m_f \!=\! 0$, this
agreement was shown up to and including $O(1/m_Q^4)$ terms in the weak
decay width in \cite{D2}, while terms up to and including
$O(m_f^2/m_Q)$ were shown to coincide with those in the OPE in
\cite{D2WA}.  In the current work we take $m_f\!=\!0$.} with the OPE was
demonstrated.

	However, our practical interest lies in the properties of
heavy hadrons with $m_Q$ lying in the intermediate domain,
specifically for $m_Q$ one order of magnitude larger than $\beta$.
The convergence of the $1/m_Q$ expansion in this case is too difficult
to quantify analytically.  This is just the situation where the
numerical computations are best employed.

	Therefore, an important element of the analysis is to check
the accuracy of the numerical computations of both the heavy hadron
masses and wavefunctions at different values of $m_Q$.  To this end,
we compute the masses and certain moments for the ground and first
excited states, and compare them to the analytic $1/m_Q$ expansion.
In general, the terms in the $1/m_Q$ expansion depend on a number of
expectation values in the static limit, like the kinetic one
$\mu_\pi^2\!=\!\aver{\bar{Q} (i\vec{D}\,)^2 Q}$, etc.  However, one can
show \cite{burkur} that the parameters appearing here through high
order in $1/m_Q$ can be expressed in terms of just the asymptotic
value $\La\!=\!M_{H_Q}\!-m_Q$ and the corresponding decay constant.
These quantities are the ones most accessible to numerical evaluation;
in particular, $\La$ is expected to be the most accurately determined
quantity.

	Our main computations refer to the case of
$m_\sp\!=\!0.56\beta$, as chosen in Ref.~\cite{GL1}.  It corresponds
(see Sec.~\ref{dual}) to a mass of the strange quark in QCD.  The
choice of a noticeable light quark mass may be motivated by an attempt
to mimic the effect of the transverse gluons absent in $D\!=\!2$,
which in a certain respect supply some effective mass to the light
quark.  Clearly, this can be only a rather crude approximation, since
the bare quark mass breaks chiral invariance.  One can suppose,
nevertheless, that this side effect is not too important for our
purposes.  The chiral symmetry is spontaneously broken anyway, and the
presence of a massless versus a massive pion does not seem to be
essential for the range of problems we address here.  On the other
hand, the effect of the transverse degrees of freedom is known to
soften the $x \to 1$ singularity of the heavy quark distribution
function \cite{burk,motion,bsg}, similar to the impact of the light
quark mass in the 't~Hooft model.  The behavior of the distribution
function affects the inclusive decays of the heavy quarks in an
essential way.

	We also present some results for $m_\sp \!=\! 0.26\beta$,
partly to explore light quark dependences of matrix elements and
partly to investigate the beginnings of failure of the numerical
solutions as $m_{\rm sp} \to 0$.  The number $N$ of Multhopp modes
used is $500$; we considered smaller $N$ as well to study this
dependence, but since the behavior was found to be stable, we do not
dwell on it further here.

	The masses of heavy hadrons obey \cite{D2,D2WA,optical}
\begin{equation}
M_{H_Q}\!-\!m_Q\,=\, \La +
\frac{\mu_\pi^2\!-\!\beta^2}{2m_Q}  +
\frac{\rho_D^3 \!-\!\rho_{\pi\pi}^3}{4m_Q^2} \;+\;
O\!\left(\frac{\beta^4}{m_Q^3}\right)
\,.
\label{50}
\end{equation}
where
\begin{eqnarray}
\mu_\pi^2 & = & \aver{\bar{Q} (i\vec{D}\,)^2 Q}\,,\;\; \rho_D^3 \, =
\, -\frac{1}{2} \aver{\bar{Q} (\vec{D} \! \cdot \! \vec{E}) Q}\,,
\nonumber \\ \rho_{\pi\pi}^3 & = & -\frac{1}{2}
\aver{iT\{\bar{Q}(i\vec{D}\,)^2 Q(x), \bar{Q}(i\vec{D}\,)^2
Q(0)\}}_{q=0}\,,
\label{npert}
\end{eqnarray}
and these expectation values refer to the infinite mass limit.  In the
$\rho_{\pi\pi}^3$ expression, $q$ is the momentum variable conjugate
to $x$, and diagonal transitions within the correlator have been
removed.

	Since $\La$ in QCD traditionally denotes the mass difference
between a ground-state pseudoscalar meson and its corresponding heavy
quark in the large $m_Q$ limit [as it is defined in Eq.~(\ref{50})],
and we need it for a number of the excited states $H_Q^{(n)}$ as well,
we assign the notation
\begin{equation}
\La^{(n)} \equiv \epsilon^{(n)} ,
\end{equation}
and use $\epsilon$ and $\La$ throughout the paper on equal footing.
Equations (\ref{50})--(\ref{58}) hold for each state $H_Q^{(n)}$ with
$n=0,1,\ldots$, so that an implicit superscript $(n)$ is to be
understood in these expressions.

	According to Ref.~\cite{burkur}, the following relations hold
in the 't~Hooft model:
\begin{equation}
\mu_\pi^2= \frac{\La^2\!-\!m^2\!+\!\beta^2}{3}\;,\qquad
\rho_D^3= \frac{\beta^2 F^2}{4}\;,\qquad
\rho_{\pi\pi}^3 = \frac{1}{36} \left[ 8\La
(\La^2\!-\!m^2\!+\!\beta^2) + 3 \beta^2 F^2 \right] \;.
\label{52}
\end{equation}
Here $F$ is the scaled decay constant in the heavy quark limit, i.e.,
\begin{equation}
F^{(n)}\;=\; \int_0^{\infty} {\rm d}t \, \Psi_n(t) \; = \;
\lim_{m_Q\to\infty} \int_0^{m_Q} {\rm d}t \, \frac{1}{\sqrt{m_Q}}
\, \phi_n \left( 1 - \frac{t}{m_Q} \right) \;=\;
\lim_{m_Q\to\infty} c_n \, \sqrt{m_Q}  \;,
\label{112}
\end{equation}
where the superscript is suppressed if there is no ambiguity,
\begin{equation} \label{cn}
c_n = \int_0^{1} {\rm d}x \: \varphi_n(x) \;,
\end{equation}
and the exact relation between $c_n$ and the decay constant of the
$n$th excitation is given in Eq.~(\ref{dec}).  In the heavy quark
limit one has
\begin{equation}
\La=m_Q \aver{1\!-\!x}\;, \qquad
\mu_\pi^2= m_Q^2\left(\aver{x^2}\!-\!\aver{x}^2\right)\;,
\label{54}
\end{equation}
but there are $O(1/m_Q)$ corrections to these relations.  For further
applications to the decay widths we also consider the scalar
expectation value \cite{D2}
\begin{equation}
\frac{1}{2M_{H_Q}}\,
\aver{\bar{Q}Q}\; = \; \frac{m_Q}{M_{H_Q}}\, \aver{\frac{1}{x}}\;.
\label{56}
\end{equation}
Then the following expansions hold:
\bea
\nonumber
\label{58}
 \sqrt{m_Q} \, c_n &=& \left( 1 -
\frac{2[2\epsilon^{(n)} \!-\!m(-1)^n]}{3m_Q} \right) F^{(n)} + O
\left(\frac{\beta^{5/2}}{m_Q^2}\right) , \\ \nonumber
m_Q \aver{1\!-\!x} &=& \La -\frac{\La^2\!+\!\mu_\pi^2}{m_Q} +
\frac{4\La(6\La^2\!+\!8\mu_\pi^2 \!+\!3\beta^2)\!+\!
\beta^2 F^2}{24m_Q^2} \,+\,
O\left(\frac{\beta^4}{m_Q^3}\right) , \\
m_Q^2\left(\aver{x^2}\!-\!\aver{x}^2\right) &=&\mu_\pi^2\,-\,
\frac{1}{3m_Q} \left(8 \La \mu_\pi^2 +\beta^2 F^2\right)\;
 + \;
O\left(\frac{\beta^4}{m_Q^2}\right)\; ,
\nonumber \\
\frac{m_Q}{M_{H_Q}}\, \aver{\frac{1}{x}} &=&
1-\frac{\mu_\pi^2\!-\!\beta^2}{2m_Q^2} -
\frac{\rho_D^3\!-\!\rho_{\pi\pi}^3}{2m_Q^3} \;+\;
O\left(\frac{\beta^4}{m_Q^4}\right) \;.
\eea

	We note that values of $\mu_\pi^2$, $\rho_D^3$, or
$\rho_{\pi\pi}^3$ determined from the expansions Eqs.~(\ref{58})
suffer degraded numerical accuracy compared to those taken directly
from Eqs.~(\ref{52}) since $\La$ and $F$ are determined from more
stable expansions (in particular, they do not depend upon close
numerical cancellations).  Therefore, we use Eqs.~(\ref{52}) as
primary information and relegate Eqs.~(\ref{58}) to numerical checks.
Our method of determining $\La$ from the $M_{H_Q}\!-m_Q$ expression,
designed to minimize the influence of potentially large uncertainties
at large $m_Q$, is described in Appendix~\ref{calc}.

\begin{figure}
  \begin{centering}
  \def\epsfsize#1#2{1.33#2}
  \hfil\epsfbox{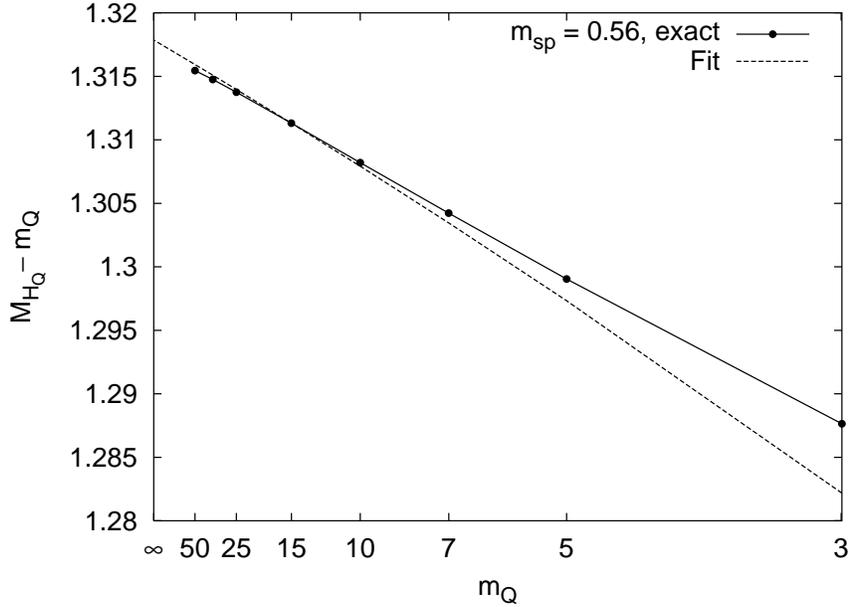}\hfil\hfill

\caption{Values of the ground-state energy $M_{H_Q}\!-m_Q$ versus
$m_Q$ for $m\!=\!0.56 \beta$ determined through direct numerical
calculation (solid line); fit of the mass expansion Eq.~(\ref{50}) to
$O(1/m_Q^2)$, using the relations Eq.~(\ref{52}) and the approach
described in Appendix~\ref{calc} (dashed line).}

\label{m56}

\end{centering}
\end{figure}

\begin{figure}
  \begin{centering}
  \def\epsfsize#1#2{1.33#2}
  \hfil\epsfbox{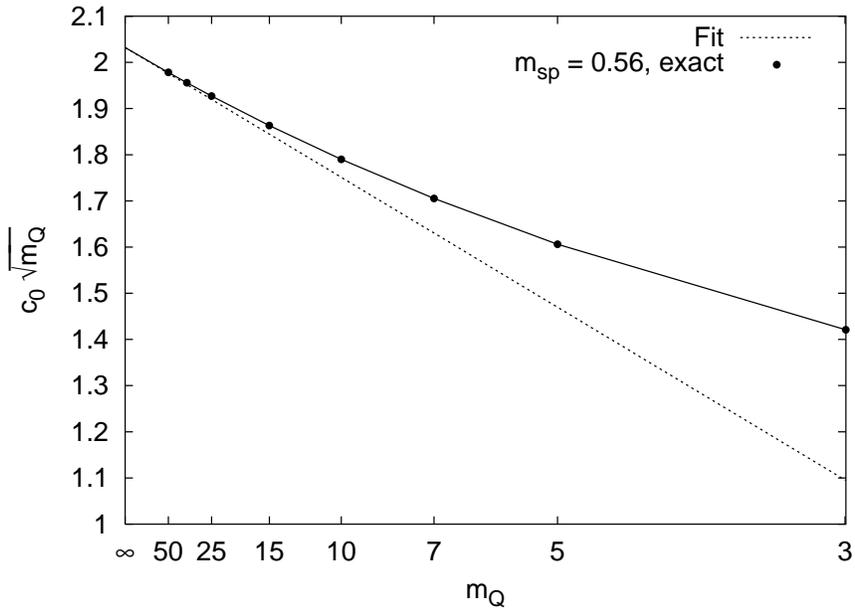}\hfil\hfill

\caption{Values of the ground-state integral $c_0 \, \sqrt{m_Q}$ for
$m \!=\! 0.56 \beta$ (solid line).  $c_0$ is related to the decay 
constant via Eq.~(\ref{dec}).  The dashed line represents an 
approximation using the first expansion in Eq.~(\ref{58}), good to 
$O(1/m_Q)$, and the asymptotic value $F$ from a polynomial fit in 
$1/m_Q$ to the points on the solid line.}

\label{dec_const}

\end{centering}
\end{figure}

\begin{figure}
  \begin{centering}
  \def\epsfsize#1#2{1.40#2}
  \hfil\epsfbox{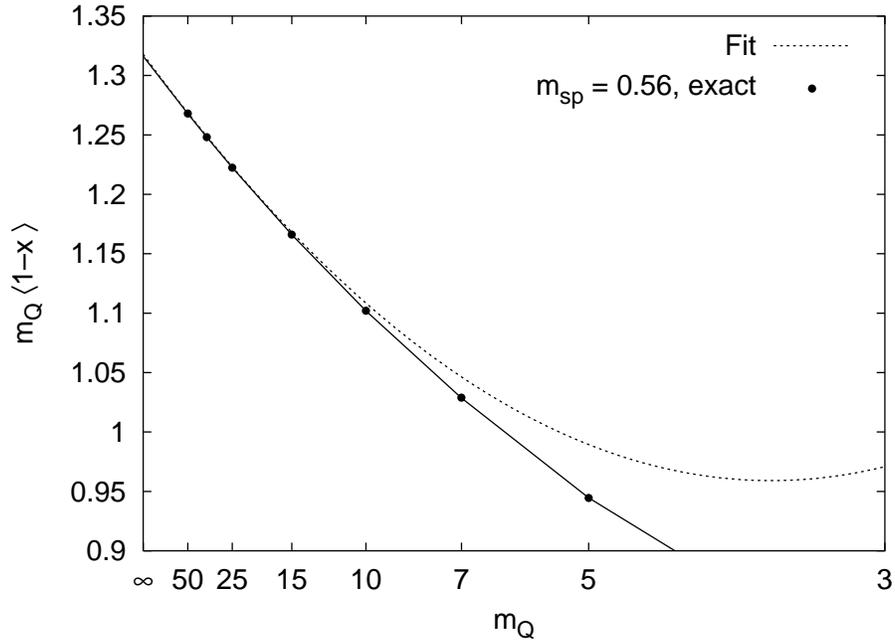}\hfil\hfill

\caption{Values of the ground-state matrix element $m_Q \langle 1\!-\!x
\rangle$ for $m \!=\! 0.56 \beta$ (solid line).  The dashed line
represents an approximation using the second expansion in
Eq.~(\ref{58}), good to $O(1/m_Q^2)$.}

\label{M1_x}

\end{centering}
\end{figure}

\begin{figure}
  \begin{centering}
  \def\epsfsize#1#2{1.40#2}
  \hfil\epsfbox{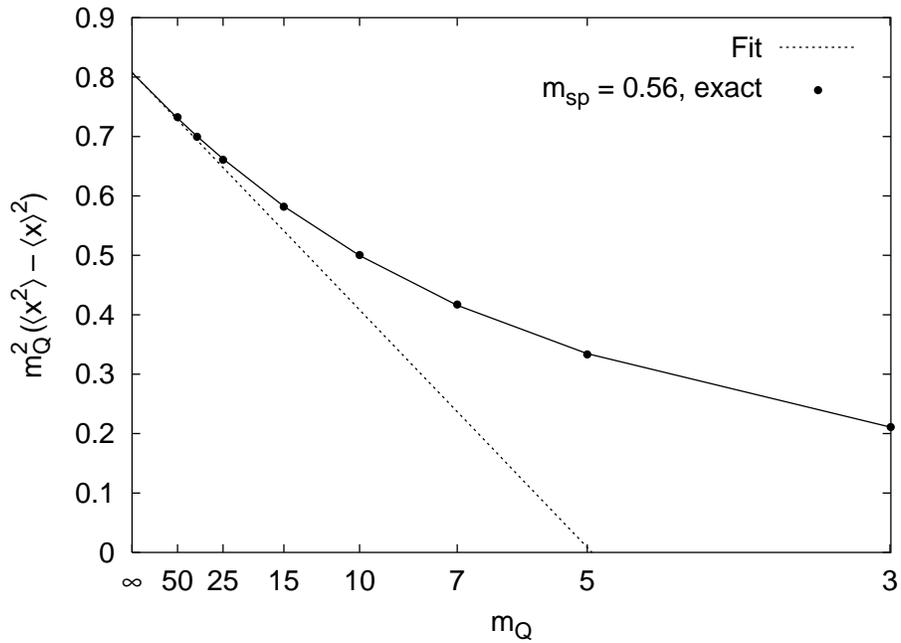}\hfil\hfill

\caption{Values of the ground-state matrix element $m_Q^2 (\langle x^2
\rangle -\! \langle x \rangle^2)$ for $m \!=\! 0.56 \beta$ (solid
line).  The dashed line represents an approximation using the third
expansion in Eq.~(\ref{58}), good to $O(1/m_Q)$.}

\label{Mx2}

\end{centering}
\end{figure}

\begin{figure}
  \begin{centering}
  \def\epsfsize#1#2{1.46#2}
  \hfil\epsfbox{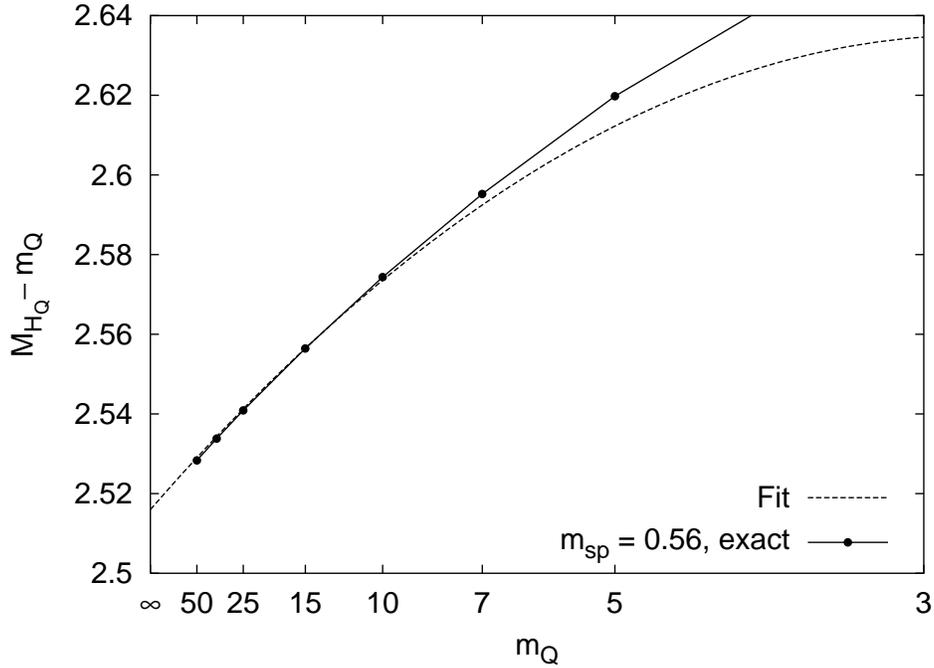}\hfil\hfill

\caption{Same as in Fig.~\ref{m56}, except for the first excited
state.}

\label{m56P}

\end{centering}
\end{figure}

\begin{figure}
  \begin{centering}
  \def\epsfsize#1#2{1.46#2}
  \hfil\epsfbox{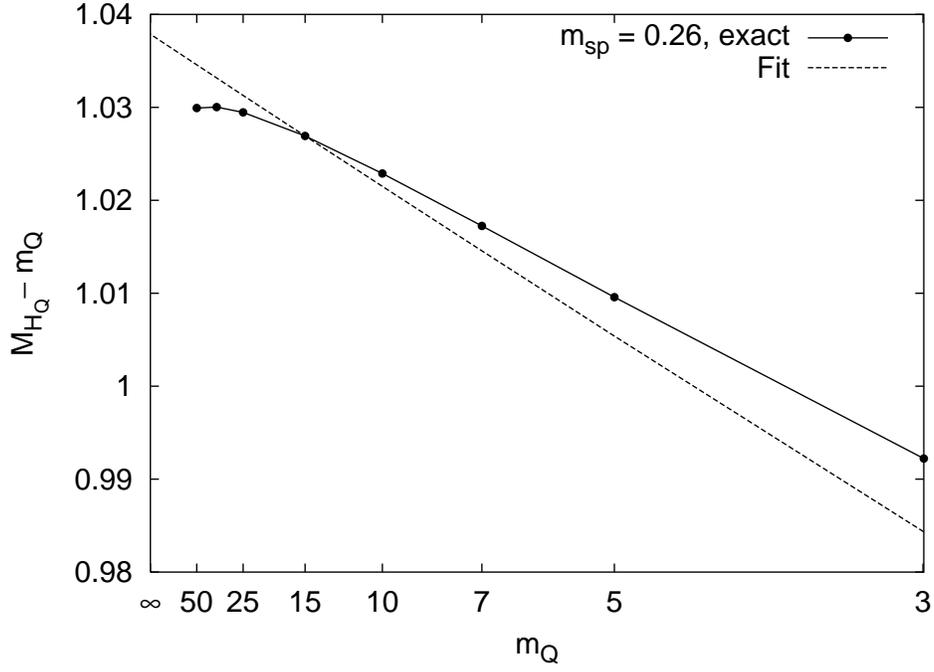}\hfil\hfill

\caption{Same as in Fig.~\ref{m56}, except $m \!=\! 0.26 \beta$.}

\label{m26}

\end{centering}
\end{figure}

	Values of $M_{H_Q}\!-m_Q$ and and the averages in
Eqs.~(\ref{58}) as functions of $m_Q$ from $m \!=\! 0.56\beta$ to
$50\beta$ are presented in Table~\ref{t1} for both the ground and
first excited states.  Similar results for just the ground state with
$m \!=\!  0.26\beta$ are presented in Table~\ref{t2}.  Based upon the
10 data points presented in Table~\ref{t1} for the ground state, one
may fit to a polynomial in $1/m_Q$, obtaining
\begin{eqnarray}
\frac{1}{\beta} (M_{H_Q} \!-\! m_Q) & = &
1.317 - 0.086 \frac{\beta}{m_Q}
- 0.050 \frac{\beta^2}{m_Q^2} + O \left( \frac{\beta^3}{m_Q^3} \right)
, \nonumber \\
c_0 \, \sqrt{\frac{m_Q}{\beta}} & = & 2.032 - 2.775 \frac{\beta}{m_Q}
+ O \left( \frac{\beta^2}{m_Q^2} \right) , \nonumber \\
\frac{m_Q}{\beta} \aver{1\!-\!x} & = & 1.316 - 2.491 \frac{\beta}{m_Q} +
3.789 \frac{\beta^2}{m_Q^2} + O \left( \frac{\beta^3}{m_Q^3} \right) ,
\nonumber \\
\frac{m_Q^2}{\beta^2} \left(\aver{x^2}\!-\!\aver{x}^2\right) & =
& 0.8074
- 4.050 \frac{\beta}{m_Q} + O \left( \frac{\beta^2}{m_Q^2} \right) ,
 \nonumber \\
\frac{m_Q}{M_{H_Q}} \, \aver{\frac{1}{x}} & = & 1 + 0.099
\frac{\beta^2}{m_Q^2} - 0.044 \frac{\beta^3}{m_Q^3} + O \left(
\frac{\beta^4}{m_Q^4} \right) .
\end{eqnarray}
The corresponding expressions using the approach of
Appendix~\ref{calc} (neglecting the one for $M_{H_Q}\!-m_Q$, which is
used as input and hence is identical through $O(\beta^2/m_Q^2)$) read
\begin{eqnarray} \label{1mfits}
c_0 \, \sqrt{\frac{m_Q}{\beta}} & = & 2.035 - 2.816 \frac{\beta}{m_Q}
+ O \left( \frac{\beta^2}{m_Q^2} \right) , \nonumber \\
\frac{m_Q}{\beta} \aver{1\!-\!x} & = & 1.318 - 2.544 \frac{\beta}{m_Q} +
4.512 \frac{\beta^2}{m_Q^2} + O \left( \frac{\beta^3}{m_Q^3} \right) ,
\nonumber \\
\frac{m_Q^2}{\beta^2} \left(\aver{x^2}\!-\!\aver{x}^2\right) & =
& 0.8078
- 3.996 \frac{\beta}{m_Q} + O \left( \frac{\beta^2}{m_Q^2} \right) ,
 \nonumber \\
\frac{m_Q}{M_{H_Q}} \, \aver{\frac{1}{x}} & = & 1 + 0.096
\frac{\beta^2}{m_Q^2} - 0.066 \frac{\beta^3}{m_Q^3} + O \left(
\frac{\beta^4}{m_Q^4} \right) .
\end{eqnarray}
This agreement between the two approaches is quite excellent and is
exhibited in Figs.~\ref{m56}--\ref{Mx2} for $M_{H_Q}\!-m_Q$ and the
quantities in Eqs.~(\ref{58}); in general, the exact results are
presented as points on a solid line, while each fit using
Eqs.~(\ref{1mfits}) is presented as a dashed line.  In Fig.~\ref{m56P}
the analogous expression $M_{H_Q}\!-m_Q$ for the $m\!=\!0.56\beta$
first excited state is presented, while Fig.~\ref{m26} uses the same
methods and values from Table~\ref{t2} to present $M_{H_Q}\!-m_Q$ for
the $m\!=\!0.26\beta$ ground state.  In Fig.~\ref{m56} and especially
in Fig.~\ref{m26}, the quality of numerical results is seen (as
expected) to begin breaking down at large $m_Q$ and small $m$, since
$N\!=\!500$ is fixed.  We conclude that the numerical routine we rely
upon is sufficiently accurate for $N\!=\!500$ up to $m_Q \approx (25
\div 30) \beta$.  The critical value of $m_Q$ also depends, however,
on the meson's light quark mass, decreasing for small $m$.  This is
expected since at small $m$ the sharpness of the wavefunction as $x\to
1$ becomes stronger, and more Multhopp functions are required to
approximate it: Each Multhopp function vanishes like $\sqrt{1\!-\!x}$.
Likewise, the required $N$ increases for the excited states.  Still,
one can check that it is possible to go as high as $m_Q \!=\! 15
\beta$ even for $m$ as small as $0.1 \beta$.

	It turns out that a numerically significant cancellation
occurs in the value of $\mu_\pi^2\!-\!\beta^2$ in $1/m_Q^2$
corrections and, in particular, at the $1/m_Q^3$ level between
$\rho_D^3$ and $\rho_{\pi\pi}^3$, for the ground state just around our
primary value $m \!=\! 0.56\beta$.  Such a numerical suppression of the
power corrections is accidental and does not occur for the excited
states, nor for $m \!=\! 0.26\beta$.

	Let us note that the expectation value of the {\em
light-quark\/} scalar density in the heavy meson turns out very close
to unity for the ground state, which may be seen by taking $m_Q
\leftrightarrow m$ and $x \leftrightarrow 1\!-\!x$ in Eq.~(\ref{56})
and referring to Table~\ref{light}; this is a characteristic feature
of a nonrelativistic (with respect to the light quark) bound-state
system.  It implies an almost simple additive dependence of $\La$ on
the light quark mass $m$,
\begin{equation}
\La \;\simeq \; \La _{\vert{m=0}}\,+\,m\;,
\label{66}
\end{equation}
and indeed one can verify this feature by comparing $M_{H_Q}\!-m_Q$
values between Table~\ref{t1} ($m\!=\!0.56\beta$) and Table~\ref{t2}
($m\!=\!0.26\beta$).  While this pattern is expected when the spectator
quark is heavy, it a priori needs not hold when it is light.  This
supports the naive expectation that the chiral symmetry breaking may
lead to a description in some aspects resembling the nonrelativistic
constituent quark model.  The above expectation value, however,
decreases for the excited states, as expected from such a picture.

	Drawing semi-quantitative conclusions for QCD requires a
translation rule between the mass parameters in the two theories, that
is, an estimate of the value of $\beta$ in $\GeV$.  Different
dimensionful quantities can be taken as the yardstick; since the
theories are not identical, this translation rule must be introduced
with some care.  As follows from the heavy quark sum rules, the
physics of duality in the decay widths of heavy flavors crucially
depends on the properties of the lowest excited heavy-quark states, in
particular the $P$-wave excitations with opposite parity to the
ground-state multiplet.  It will become evident from the next section
that they are of primary importance for the $1/m_Q$ expansion of
static properties as well.  Therefore, we choose the mass difference
between the lowest parity-even ($P$-wave) state and the parity-odd
ground-state meson to gauge the translation between the mass scales.

	In the 't~Hooft model the mass difference
$\epsilon_1\!-\!\epsilon_0$ for light spectators amounts to about
$1.3\beta$.  Real charm spectroscopy suggests that the first $P$-wave
excitations are between $400$ and $500 \MeV$ above the ground state.
Taking the larger value for sake of illustration, we arrive at the
estimate
\begin{equation}
\beta\;\approx \; 400 \MeV \; ,
\label{68}
\end{equation}
which is adopted in our analysis.  This falls rather close to the
estimate of Ref.~\cite{burkswan}, which relied on a quite different
type of effects in the light-quark systems.

	Assuming a value for the ``bare'' $b$ quark mass in QCD
(normalized at the appropriate scale $\gsim m_b$) of about $4\GeV$, we
conclude that mesons with quarks of masses
\begin{equation}
m_b \approx 10\beta \; , \ \
m_c \approx (2.5\div 3.5)\beta \; ,
\label{80}
\end{equation}
represent in the 't~Hooft model the actual beauty and charm mesons.

	The value of $\La\simeq 1.3\beta \approx 500\MeV$ seems to be
in a reasonable correspondence with the size of this difference in QCD
when it is normalized at a low hadronic scale, $\La(1\GeV) \approx
(600\pm 60)\MeV$ \cite{mymass}.

	It should be noted, however, that the kinetic expectation
value in the 't~Hooft model turns out to be rather small, $\mu_\pi^2
\simeq 0.8\beta^2 \approx 0.12\GeV^2$.  This is not surprising, since
the chromomagnetic field is absent in two dimensions, while it was
shown \cite{optical,volpp,rev} to be crucial in the real case.
Indeed, the comparison is better justified for the difference
$\mu_\pi^2\!-\!\mu_G^2$ in actual QCD versus the value of $\mu_\pi^2$
in the 't~Hooft model.  These questions were discussed in detail in
Ref.~\cite{varenna}, and can be easily understood using the sum rule
representation.  Due to the absence of spin in two dimensions, there
is no difference between the would-be spin-$1/2$ and spin-$3/2$ light
degrees of freedom.  In particular, labeling the ``oscillator
strengths'' $\tau$ defined in the next section by spin rather than
excitation number, $\tau_{1/2}\!=\!\tau_{3/2}$ and
$\epsilon_{1/2}\!=\!\epsilon_{3/2}$ effectively hold.  Then the sum
\begin{equation}
\mu_\pi^2 \;=\; 3\sum_n \,\epsilon_n^2|\tau_{1/2}|^2 \:+\:
6\sum_n \,\epsilon_n^2|\tau_{3/2}|^2 \;\to\;
9\sum_n \,\epsilon_n^2|\tau_{1/2}|^2 \; ,
\label{82}
\end{equation}
and the latter sum is just the general expression for
$\mu_\pi^2\!-\!\mu_G^2\,$:
\begin{equation}
\mu_\pi^2 \!-\!\mu_G^2\;=\; 9\sum_n \, \epsilon_n^2|\tau_{1/2}|^2 \;.
\label{84}
\end{equation}
Accepting such an identification suggested in Ref.~\cite{varenna} and
the estimate $\mu_\pi^2\!-\!\mu_G^2 \simeq (0.15\pm 0.1)\GeV^2$, we
again observe a reasonable agreement with the findings of the 't~Hooft
model.

\section{Duality in the SV Sum Rules}
\label{dual}

	A useful theoretical limit---the so-called small velocity (SV)
regime---was suggested in the mid 80's \cite{SV} as a theoretical tool
for studying semileptonic heavy quark decays.  This refers to
kinematics where both $b$ and $c$ quarks are heavy, but the energy
release is limited, so that the velocity of the final charm hadron is
small.  At large energy release the OPE for the width must converge
rapidly to the actual hadronic width.  Still, at fixed energy release
the deviations, although $1/m_Q$ suppressed, are present regardless of
the absolute values of masses.

	In the SV regime the semileptonic decays proceed either to the
ground-state charm final state, $D$ or $D^*$ (the semi-elastic
transitions), or to excited ``$P$-wave'' states of the opposite
parity.  Other decays are suppressed by higher powers of velocity, or
by heavy quark masses.

	The equality of the sum of partial decay widths and its OPE
expansion is achieved through the sum rules that relate the sums of
the $P$-wave transition probabilities, weighted with powers of the
excitation energies, to the static characteristics of the decaying
heavy hadron.  The onset of convergence of the OPE expansion for the
widths is then directly related to the pattern of saturation of the
sum rules by the lowest excitations.  If higher states contribute
significantly, they delay the onset of duality, while their absence
leads to a tight quark-hadron duality after the first $P$-wave channel
is open.

	Knowledge of degree of saturation of the heavy quark sum rules
is also important for another reason: It determines the hadronic scale
above which one can apply the perturbative treatment to compute
corrections or account for evolution of the effective operators.  The
lower this scale, the more predictive in turn is the treatment of the
nonperturbative effects in the OPE.

	A recent review of the SV sum rules can be found in
Ref.~\cite{rev} (the perturbative aspects are discussed in more detail
in Ref.~\cite{varenna}).  For most practical purposes addressed here,
one can consider the perturbative effects to be absent in the 't~Hooft
model.  In particular, the heavy quark parameters do not depend
perturbatively on the normalization point, and there is no need in the
explicit ultraviolet cutoff to introduce a normalization point.  The
sum rules we address are
\bea
\rho_k^2\!-\!\frac{1}{4} &=& \sum_n \label{oscsum1}\; \,\tau_{nk}^2 \; ,
\\ \frac{1}{2}\, \La_k\; &=& \sum_n (\epsilon_n\!-\!\epsilon_k)\;
\tau_{nk}^2 \; ,
\\ \left(\mu_\pi^2\right)_k &=& \sum_n (\epsilon_n\!-\!\epsilon_k)^2
\,\tau_{nk}^2 \; ,
\\
\left(\rho_D^3\right)_k &=& \sum_n
(\epsilon_n\!-\!\epsilon_k)^3 \,\tau_{nk}^2 \; .
\label{oscsum4}
\eea
Here $k$ and $n$ denote excitation indices for the initial and final
states, respectively (in practice only transitions from the ground
state are interesting, so we limit ourselves to $k=0$; in this case
the index $k$ is omitted).  The so-called ``oscillator strengths''
$\tau$ parameterize the transition amplitudes into the opposite-parity
states in the SV limit,
\beq
\frac{1}{2m_Q} \matel{n}{\bar{Q}\gamma_\mu Q}{k}\;=\;  \tau_{nk} \,
\epsilon_{\mu\nu} v^\nu \,+\, O(\vec{v}^{\,3}) ,
\label{102}
\eeq
where $\vec v$ is the velocity of the final state hadron.  In the
diagonal transition $\rho_k^2$ is the slope of the Isgur-Wise (IW)
function of state $\state{k}$:
\beq
\frac{1}{2m_Q} \matel{k(\vec{v})}{\bar{Q}\gamma_0 Q}{k(0)}\;=\;
 1-\rho_k^2 \frac{\vec{v}^{\,2}}{2} \,+\,
 O (\vec{v}^{\,4})\;.
\label{104}
\eeq
The expressions for $\tau_{nk}$ and $\rho_k$ in terms of the
light-cone wavefunctions are
\bea
\nonumber
\tau_{nk} & =&  \int_0^{\infty} {\rm d}t \, \Psi_n(t)\,
t\frac{{\rm d}}{{\rm d}t} \Psi_k(t) = -\lim_{m_Q\to\infty}
\int_0^{1} {\rm d}x \, \varphi_n(x)\,
(1\!-\!x)\frac{{\rm d}}{{\rm d}x} \varphi_k(x) , \\
\rho^2_{k} & =&  \int_0^{\infty} {\rm d}t \,
\left|\left(t\frac{{\rm d}}{{\rm d}t} +\frac{1}{2}\right)
\Psi_k(t)\right|^2 = \lim_{m_Q\to\infty}
\int_0^{1} {\rm d}x \, \left|\left[(1\!-\!x)\frac{{\rm d}}{{\rm d}x}
-\frac{1}{2}\right]\varphi_k(x)\right|^2\;. \nonumber \\
\label{106}
\eea
The finite-$m_Q$ corrections to the $x$-integral forms turn out to be
rather significant, leading to significant problems in precision
numerical studies.  To avoid this problem we use the analytic
expression for the inelastic amplitudes obtained in
Ref.~\cite{burkur}:
\begin{equation}
\tau_{nk} = \matel{n}{t\frac{\rm d}{{\rm d}t}}{k} \;=\;
- \frac{\beta^2}{2(\epsilon_n\!-\!\epsilon_k)^3} \, F^{(n)} F^{(k)}
\left(\frac{1\!-\! (-1)^{n-k}}{2} \right) \;,
\label{110}
\end{equation}
where $F^{(n)}$ are the asymptotic values of the decay constants $c_n$
scaled up by the factor $\sqrt{m_Q}$, as in Eq.~(\ref{112}).  The
constants $F^{(n)}$ are computed as the values of $c_n \, \sqrt{m_Q}$
at $m_Q \!=\! 15 \beta$ (see Table~\ref{t1}) augmented by the $1/m_Q$
corrections detailed in the first of Eqs.~(\ref{58}), while values of
$\epsilon_n$ are computed using the procedure described in
Appendix~\ref{calc}.

	The results of the computations for the case $m_Q \!=\! 15
\beta$, $m \!=\! 0.56 \beta$ are presented in Table~\ref{osc}.  Our
central result is a surprisingly good saturation of the sum rules: The
first $(n\!=\!1)$ excitation generates $99.4\%$ of $\rho^2$, $98.5\%$
of $\La$, $96\%$ of $\mu_\pi^2$, and even $91\%$ of $\rho_D^3$.  The
rest is almost completely saturated by the second $P$-wave state
($n\!=\!3$), where the cumulative values for the same quantities read
$99.92\%$, $99.73\%$, $99.1\%$, and $96.7\%$, respectively.

	In terms of absolute numbers, the sum rules
Eqs.~(\ref{oscsum1})--(\ref{oscsum4}) would give $\rho^2\!-\!1/4 \!=\!
0.529$, $\La \!=\! 1.278 \beta$, $\mu_\pi^2 \!=\! 0.782 \beta^2$, and
$\rho_D^3 \!=\! 0.99 \beta^3$, the last of which gives $F^{(0)} \!=\!
1.99 \sqrt{\beta}$, in fine agreement with the values obtained from
the values obtained in the previous section via the methods described
in Appendix~\ref{calc}.  The few-percent discrepancy corresponds to
the accuracy in determinations of squared decay constants.

	The level of saturation by the lowest open channels is
extraordinary.  The explicit reason for such a perfect saturation of
the sum rules involving even rather high, $\sim \epsilon^3$ powers of
the excitation energy can be read off Eq.~(\ref{110})---$\,\tau$'s are
inversely proportional to the third power of the excitation energy.
With the asymptotics $\epsilon_n \sim \sqrt{n}$, $F^{(n)} \sim
n^{-1/4}$, the first excitation energy $\epsilon_1\!-\!\epsilon_0$ is
notably smaller than the next one $\epsilon_3\!-\!\epsilon_0$ including
three energy gaps.  The general peculiarity of the 't~Hooft model
leading to such a saturation is not understood completely.

	With this pattern of saturation of the SV sum rules for the
ground-state meson, one expects an early onset of the accurate duality
for the inclusive widths in the $b\to c$ transitions, only slightly
above the threshold of the first excitation.  Demonstrating this
result through direct evaluation of the decay widths is one of the
purposes of the next section.

\section{Local Duality in the Decay Widths}
\label{wid}

	The semileptonic widths described in this work were considered
in detail in Ref.~\cite{D2}.  Here we recapitulate a few basic points.
The weak decay Lagrangian is
\begin{equation}
{\cal L}_{\rm weak} \,=\, -\frac{G}{\sqrt{2}}\,(\bar c \gamma_\mu b)
\,(\bar{e}\gamma^\mu \nu )\;.
\label{120}
\end{equation}
In terms of the previous notation, $Q \to b$, $q \to c$ (or, later in
this section, $u$), and $H_Q \to B$.  The key property of all
$D\!=\!2$ vectorlike currents is that for $m_e \!=\!m_\nu\!=\!0$, the
invariant mass $q^2$ of the lepton pair is always zero.  For all
computational purposes decays into this massless lepton pair are
equivalent to decays into a single massless pseudoscalar particle
$\phi$ weakly coupled to quarks according to
\begin{equation}
{\tilde{\cal L}}_{\rm weak} \,=\,-\frac{G}{\sqrt{2\pi}}\;
\bar c \gamma_\mu b \: \epsilon^{\mu\nu} \,\partial_\nu \phi \; .
\label{122}
\end{equation}

	Several arguments favor our choice of a vectorlike weak decay
interaction in the 't~Hooft model.  One is of course the simplicity of
Eq.~(\ref{122}).  Another is that for $q^2\!=\!0$ some difficult
problems of renormalization are absent, as we now discuss.  The central
problem in applying the OPE in practice is disentangling perturbative
and nonperturbative effects.  More precisely, this refers to the
separation of short-distance effects attributed to the coefficient
functions from long-distance effects residing in the matrix elements
of the effective heavy-quark operators.

	The perturbative corrections, for example those that
renormalize the weak quark current, are generally rather nontrivial,
even in the 't~Hooft model.  However, according to the
nonrenormalization theorem of Ref.~\cite{D2}, such vertex corrections
are absent from the decays with $q^2\!=\!0$.  This allows one to
isolate the problem of renormalization of the underlying current from
the question of interest in our study: possible deviations of the full
decay widths due to the presence of thresholds in the production of
the hadronic resonances.

	In reality, from the OPE viewpoint some short-distance
corrections still remain even in this special kinematic region due to
the high-momentum tails in the meson wavefunctions.  These tails come
from the hard gluon exchanges between the constituents.  In principle,
these ``hard'' components can also be separated from the ``soft''
bound-state dynamics explicitly.  However, in practice this is not
necessary: These effects are completely contained in the meson
wavefunctions.

	Another advantage of vectorlike currents is apparent when one
notes that the $D\!=\!2$ three-body ``semileptonic'' phase space
diverges logarithmically for massless leptons.  Explicitly, for the
decay $M \to m \!+\! m_\ell \!+\! m_\ell$ (equal lepton masses are
assumed to render the expressions simpler), the three-body phase space
turns out to be
\begin{eqnarray}
\Phi_3 ( M; m, m_\ell, m_\ell ) & = & \frac{1}{4 \pi^3 (M\!-\!m)
\sqrt{(M\!+\!m)^2\!-\!4m_\ell^2}}
\nonumber
\\ & & \times K \left[
\frac{(M\!+\!m)^2
\left[ (M\!-\!m)^2 \!-\!4m_\ell^2 \right]}{(M\!-\!m)^2 \left[
(M\!+\!m)^2 \!-\!4m_\ell^2 \right]} \right],
\end{eqnarray}
where $K$ is the complete elliptic integral of the first kind.  As
$m_\ell \to m$, one regains Eqs.~(4.1)--(4.3) of \cite{GL1}, while as
$m_\ell \to 0$, the argument of the elliptic integral goes to unity,
and $K(1\!-\!\epsilon) \to \ln (8/\epsilon)/2$.  This is a manifestation
of the logarithmic infrared divergence of the massless scalar Green
function at large distance in $D\!=\!2$.  A detailed calculation shows
that the vector nature of the weak coupling regularizes the phase
space integral, preventing the partonic rate from diverging in the
limit of massless leptons.  Furthermore, as discussed in the
Introduction, this also removes the $1/\mvec{p}$ singularity in the
threshold behavior for hadronic two-body decays.

	As a final advantage of vectorlike currents and the special
kinematic point $q^2\!=\!0$, note that at $q^2\!=\!0$ the $B\to D^{(n)}$
transition amplitudes are directly expressed in terms of the overlap
between the initial and the final wavefunctions:
\begin{equation}
q_\mu\: \frac{1}{2M_B} \langle n|\epsilon^{\mu\nu} J_\nu |B \rangle
\;=\;
-q_z\, \int_0^1 \, {\rm d}x \: \varphi_n(x) \varphi_B(x)
\;,
\label{123}
\end{equation}
where $q_z = -\mvec{p} = -(M_B^2 -\! M_n^2)/2M_B$, so that the
partial decay width for $B\to D^{(n)} \, \ell \bar\nu$ is given by
\begin{equation}
\Gamma_n =  \frac{G^2}{4\pi } \cdot
\frac{M_B^2 \!-\! M_n^2}{M_B}\,
\left| \int _0^1 {\rm d}x \, \varphi_n(x)\varphi_B(x) \right|^2
\,\theta(M_B-M_n)
\;.
\label{124}
\end{equation}
The threshold suppression mentioned above is manifested in the
explicit factor $(M_B^2\!-\!M_n^2)$: The reciprocal of this factor in
the phase space is removed by $q_z^2$ from the matrix element.  It is
also possible to derive this result directly using the methods of
Ref.~\cite{GL1}; note, however, that these expressions are much
simpler than those of Ref.~\cite{GL1}, because the vectorlike current
with massless leptons restricts $q^2$ to 0.  The sum of these widths
over all open channels is to be compared to the OPE prediction.  The
remarkable speed of saturation in $n$, anticipated in the last
section, is illustrated for one sample case in Table~\ref{satur}.

	Turning to the OPE, we mention one more problem associated
with an accurate understanding of local duality violation.  Apart from
the purely theoretical aspect that OPE power series are generally only
asymptotic and, thus have a formally zero radius of convergence in
$1/m_Q$, one normally has additional practical limitations.  Only a
limited number of the terms, as well as the associated expectation
values, are usually known, which places additional theoretical
uncertainties that dominate in practice at sufficiently large $m_Q$.

	This feature can be naturally incorporated in the analysis of
our concrete model.  We account completely only through terms that
scale like $1/m_Q^4$, the highest order that emerges from the OPE free
from the four-fermion operators \cite{D2}.  The rest, although
calculable in principle term-by-term in the 't~Hooft model, are taken
to represent the OPE ``tails'' discarded by the unavoidable
truncation.

	Using the sum rules of the 't~Hooft model, Ref.~\cite{D2}
established the following exact representation for the total decay
width:
\begin{equation}
\Gamma_B= \frac{G^2}{4\pi}\cdot \frac{m_b^2 \!-\!
m_c^2}{M_B}
\int _0^1 \frac{{\rm d}x}{x}
\,\varphi_B^2(x) \;- \;\sum _{M_n> M_B} \Gamma_n ,
\label{130}
\end{equation}
where $\Gamma_n$ at $M_n \!>\!M_B$ are understood as given by
Eq.~(\ref{124}) without the explicit $\theta$-function singling out
the open channels; such $\Gamma_n$ are therefore all negative.  On the
other hand, the OPE yields the result
\begin{equation}
\Gamma_B= \frac{G^2}{4\pi}\cdot \frac{m_b^2 \!-\!
m_c^2}{m_b} \left[ \frac{m_b}{M_B} \int_0^1
\frac{{\rm d}x}{x} \,
\varphi_B^2(x) \,+ \,O\left(\frac{\beta^5}{M^5}\right) \right] \; ,
\label{132}
\end{equation}
with $M$ generically denoting the OPE expansion parameter; we do not
specify here if it is $m_b$ or $m_b\!-\!m_c$, or some other combination.
It was shown in Ref.~\cite{D2} that the $\Gamma_n$ term in
Eq.~(\ref{130}) is dual to the order term in Eq.~(\ref{132}); however
we do not use this here and rather treat the latter as an intrinsic
uncertainty in the ``practical'' version of the OPE.

	Thus, our strategy is to compare the exact width
\begin{equation}
\Gamma_B =  \frac{G^2}{4\pi }
\sum_{M_n<M_B} \,\frac{M_B^2 \!-\! M_n^2}{M_B}\,
\left| \int_0^1 {\rm d}x \varphi_n(x)\varphi_B(x) \right|^2 ,
\label{140}
\end{equation}
to
\begin{equation}
\Gamma_{\rm OPE}= \frac{G^2}{4\pi} \cdot \frac{m_b^2 \!-\!
m_c^2}{m_b} \cdot \frac{m_b}{M_B} \int_0^1 \frac{{\rm d}x}{x}
\,\varphi_B^2(x)\, .
\label{141}
\end{equation}

	The expectation value $\frac{m_b}{M_B} \aver{\frac{1}{x}}$
above can either be evaluated numerically, or in the spirit of the
OPE, computed in the form of a $1/m_b$ expansion, the last of
Eqs.~(\ref{58}).  It turns out that the expansion converges very
rapidly to the exact result, so that this does not significantly
affect the observed pattern of local duality at the quantitative
level.  The Born-term partonic rate is simply given by $\Gamma_{\rm
OPE}$ with this expectation value set to unity,
\begin{equation}
\label{Born}
\Gamma_b = \frac{G^2}{4\pi} \cdot \frac{m_b^2 \!-\! m_c^2}{m_b} .
\end{equation}

	The main practical interest of these calculations lies in the
$b\!\to\! c$ width with its limited energy release $E_r$.  In general,
$E_r$ can be small either if $m_b$ is not large enough, or even at
large $m_b$ if $E_r \!=\! m_b\!-\!m_c$ (or
$m_b\!-\!m_c\!-\!\sqrt{q^2}$ if $q^2$ is nonzero) is insufficient due
to a significant $c$ quark mass.  The latter case falls into the SV
category, and the violations of duality are suppressed here even at
the maximal $q^2$ by heavy quark symmetry, as was pointed out in the
mid-80's \cite{SV}.  Therefore, one a priori expects a different
pattern in the two cases.  We try to separate the possible effects by
considering different choices for $m_b$ and $m_c$ rather than by only
taking them close to their realistic values.

	With these arguments in mind, one can expect to find
significant effects of duality violation in the cases where $1/m_c$ or
$1/m_b$ effects are important.  As suggested in Ref.~\cite{varenna},
in this case it is advantageous to fix $m_b$ close to its actual
value, and vary $m_c$ from near $m_b$ down to smaller values, changing
in this way the energy release.  At one end of the interval the local
duality is supported by the heavy quark symmetry with large quark
masses and SV kinematics, while at another end it rests on the large
energy release.

\begin{figure}
  \begin{centering}
  \def\epsfsize#1#2{1.40#2}
  \hfil\epsfbox{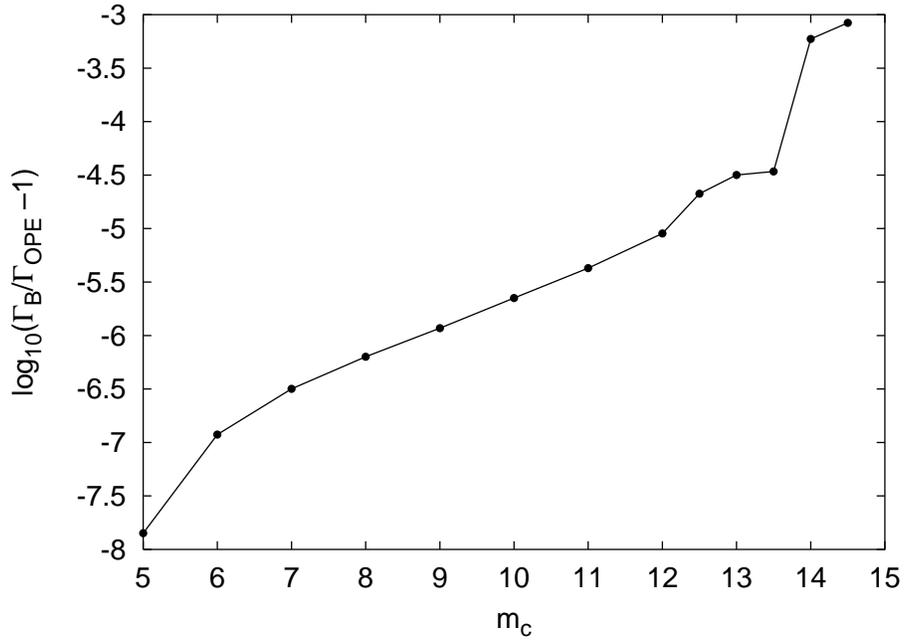}\hfil\hfill

\caption{Duality deviation between exact hadronic width $\Gamma_B$ and
$\Gamma$ determined from the OPE [Eq.~(\ref{141}) or (\ref{C132})],
good to $O(1/m_Q^4)$.  Here $m_b \!=\! 15 \beta$, $m \!=\! 0.56 \beta$, 
and $m_c$ variable.}

\label{dual15fig}

\end{centering}
\end{figure}

\begin{figure}
  \begin{centering}
  \def\epsfsize#1#2{1.40#2}
  \hfil\epsfbox{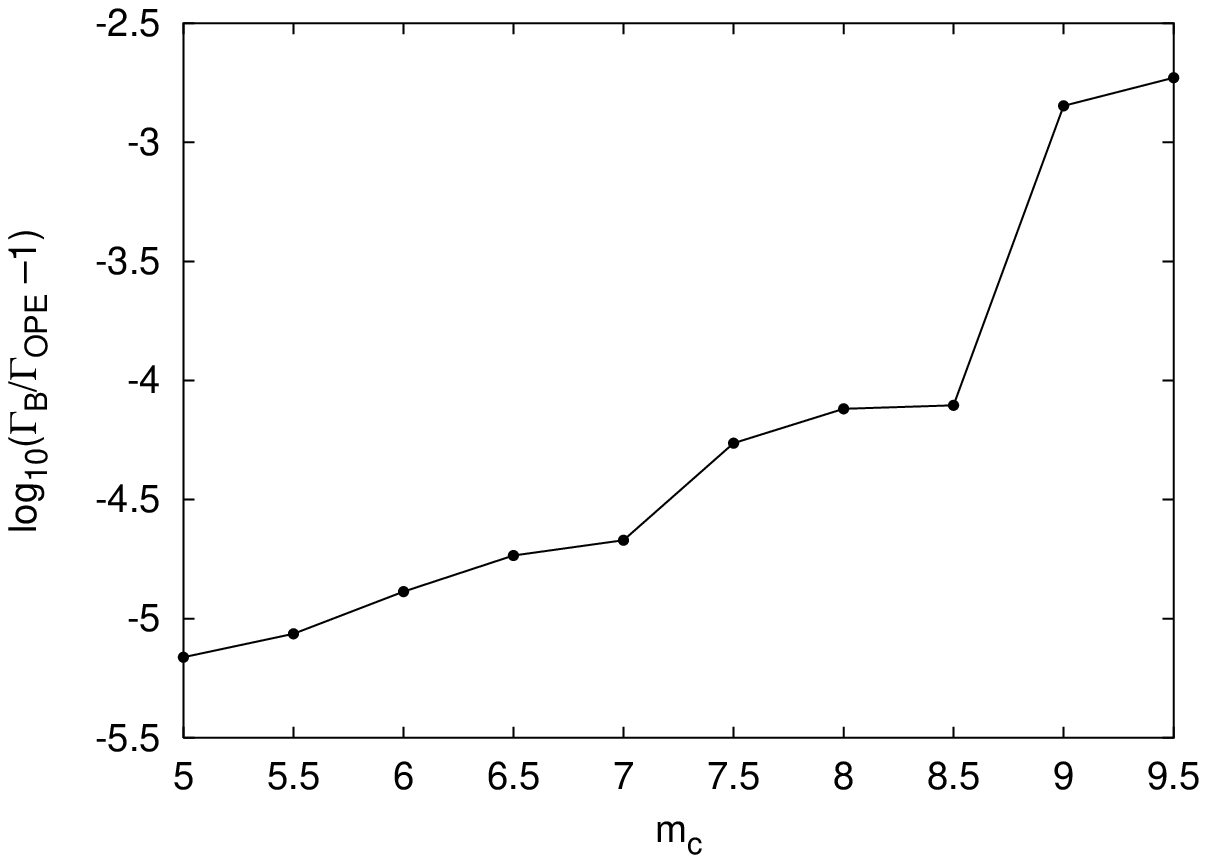}\hfil\hfill

\caption{Same as Fig.~\ref{dual15fig}, except $m_b \!=\! 10\beta$.}

\label{dual10fig}

\end{centering}
\end{figure}

	We start from the SV case when $m_b$ is fixed and large and
$m_c$ is large as well, varying the energy release by increasing $m_c$
towards $m_b$.  Since the violation of local duality is expected to be
suppressed for all values of $m_c$, high numerical accuracy is vital.
We fix $m_b\!=\!15\beta$ ($\approx 6\GeV$), and vary $m_c$ from
$5\beta$ up to $m_b$.  The results are given in Table~\ref{dual15} and
Fig.~\ref{dual15fig}.  We note that the difference between the two
widths is so small that one must plot $\ln (\Gamma_B/\Gamma_{\rm
OPE}\!-\!1)$ rather than the widths themselves.  This is expected
since the SV sum rules are very well saturated, as detailed in the
previous section---the higher thresholds are then strongly suppressed
numerically at finite energy release.  But for $m_c$ approaching
$m_b$, where they could be noticeable, the heavy quark symmetry works
efficiently since both quarks are very heavy.  In fact, the only
prominent features on the plot occur when thresholds to the first few
$D$ states of opposite parity to the ground-state $B$ meson are
crossed, for example between $m_c \!=\! 13.5$ and $14\beta$.  The
deviation is extremely small also for smaller $m_c$ where the $c$
quark velocity is rather large---yet there the energy release is
significant, and a large number of excited states (up to $18$ at $m_c
\!=\! 5 \beta \approx 2\GeV$) are produced.  Table~\ref{dual10} and
Fig.~\ref{dual10fig} show analogous results for $m_b \!=\! 10 \beta$,
$m \!=\! 0.56 \beta$.

\begin{figure}
  \begin{centering}
  \def\epsfsize#1#2{1.40#2}
  \hfil\epsfbox{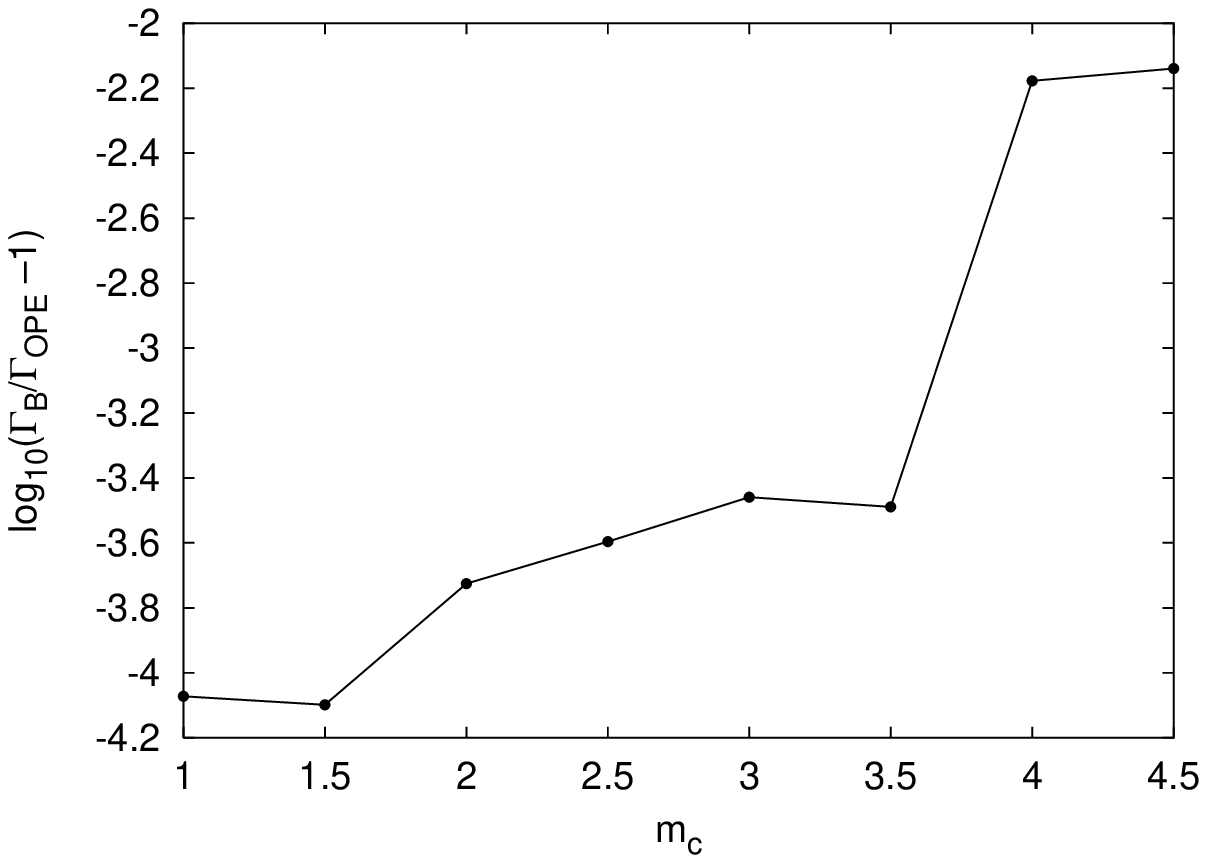}\hfil\hfill

\caption{Same as Fig.~\ref{dual15fig}, except $m_b \!=\! 5\beta$.}

\label{dual5fig}

\end{centering}
\end{figure}

	To render the duality violation more apparent, we consider
(Table~\ref{dual5} and Fig.~\ref{dual5fig}) the same decay widths for
a $b$ quark with half the mass, $m_b\!=\!5\beta\approx 2\GeV$.  Even
here the deviation is below per mill as soon as the first excitation
can appear with sufficient phase space.  The duality-violating
component at last exhibits the proper oscillating behavior (note the
decrease between $m_c \!=\! 3$ and $3.5\beta$ or 1 and $1.5\beta$), but
this effect is too small to be extracted reliably at larger energy
release where this property becomes an asymptotic rule.

	As follows from our computations, local duality is violated at
a tiny level in the $b\to c$ decays in the 't~Hooft model whenever it
is a priori meaningful to apply OPE.  A possible reason behind this
might be that for unidentified reasons the heavy quark symmetry works
for the inclusive widths too effectively, down to relatively low
masses and velocities of order $1$.  This was conjectured in the early
papers on the subject \cite{SV}.

\begin{figure}
  \begin{centering}
  \def\epsfsize#1#2{1.40#2}
  \hfil\epsfbox{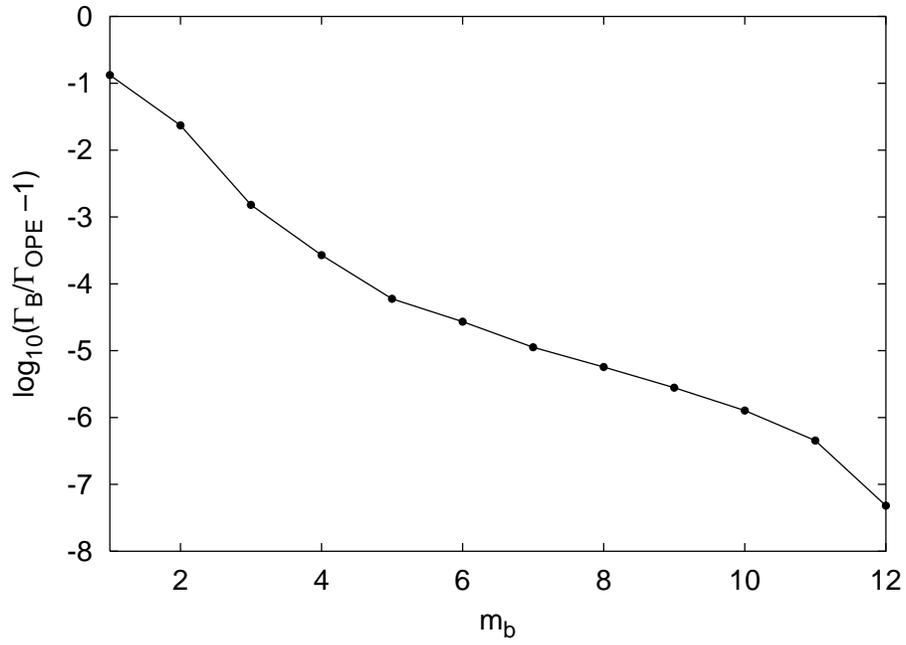}\hfil\hfill

\caption{Same as Fig.~\ref{dual15fig}, except $m_c \!=\! m \!=\! 0.56 
\beta$ and $m_b$ variable.}

\label{dual_mc56}

\end{centering}
\end{figure}

\begin{figure}
  \begin{centering}
  \def\epsfsize#1#2{1.40#2}
  \hfil\epsfbox{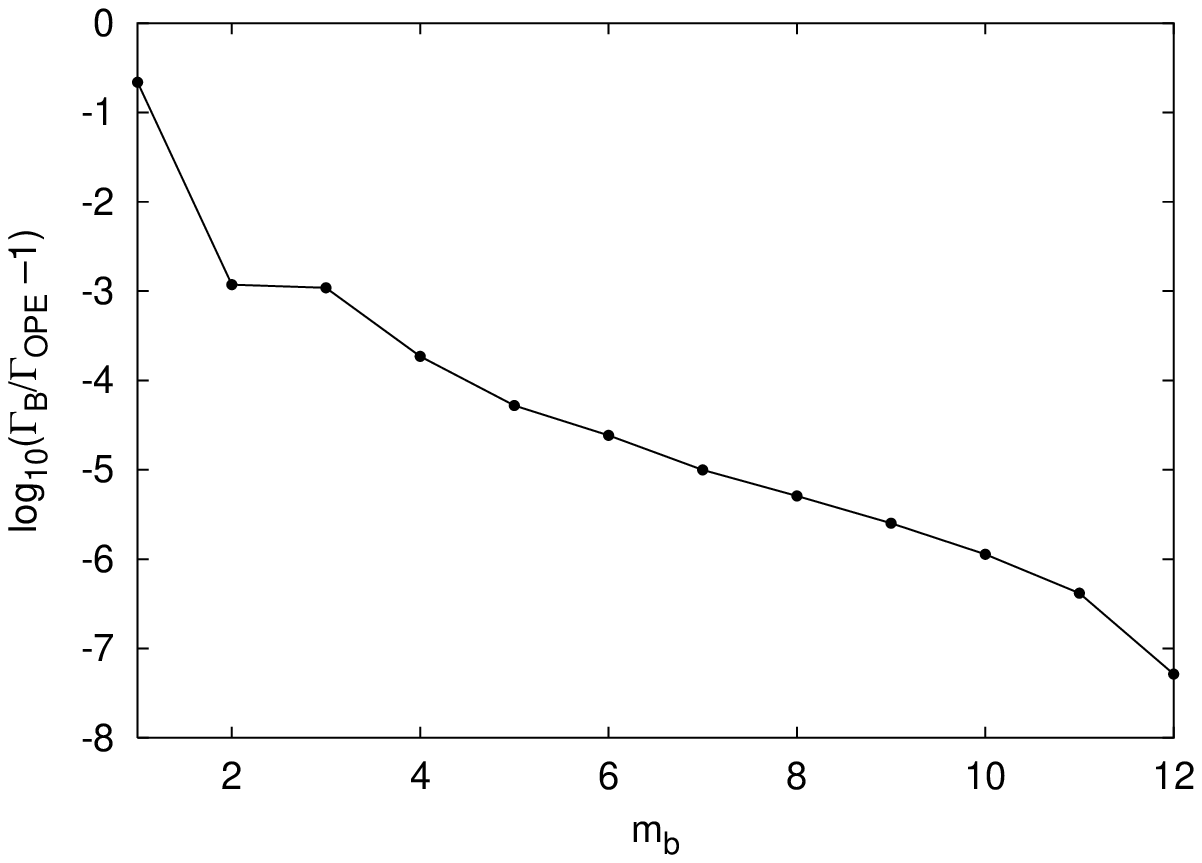}\hfil\hfill

\caption{Same as Fig.~\ref{dual_mc56}, except $m_c \!=\! m \!=\! 0.26 
\beta$.}

\label{dual_mc26}

\end{centering}
\end{figure}

	Therefore, our final attempt in the quest for the sizeable
duality violation in beauty is considering the $(b\!\to \!u)$-type
transitions, where the heavy quark symmetry per se does not constrain
the individual transition form factors.  We fix in our expressions
$m_c\! =\! m\! =\! 0.56 \beta$ or $0.26 \beta$ (but still keep the two
quarks flavor-distinguished) and vary $m_b$ from $1\beta \!\approx\!
0.4\GeV$ to $12\beta \!\approx\! 4.5\GeV$.  The results are shown in
Table~\ref{dualmc56} and Fig.~\ref{dual_mc56}, and
Table~\ref{dualmc26} and Fig.~\ref{dual_mc26}, respectively.  Although
the difference between the actual width and its OPE approximation is
larger, it still is very small and approaches a percent level for
$m_b$ as low as $2\beta \!\approx\! 0.8\GeV$.  The total decay width is
no longer saturated to such a high degree by transitions to the ground
state, especially for larger $m_b$.  Nevertheless, the duality is
amazingly well satisfied when just the first few open channels are
summed.  Again, the only prominent features in the plots appear when
crossing kinematic thresholds due to the lightest $D$ mesons of
opposite parity to the ground-state $B$.

\begin{figure}
  \begin{centering}
  \def\epsfsize#1#2{1.40#2}
  \hfil\epsfbox{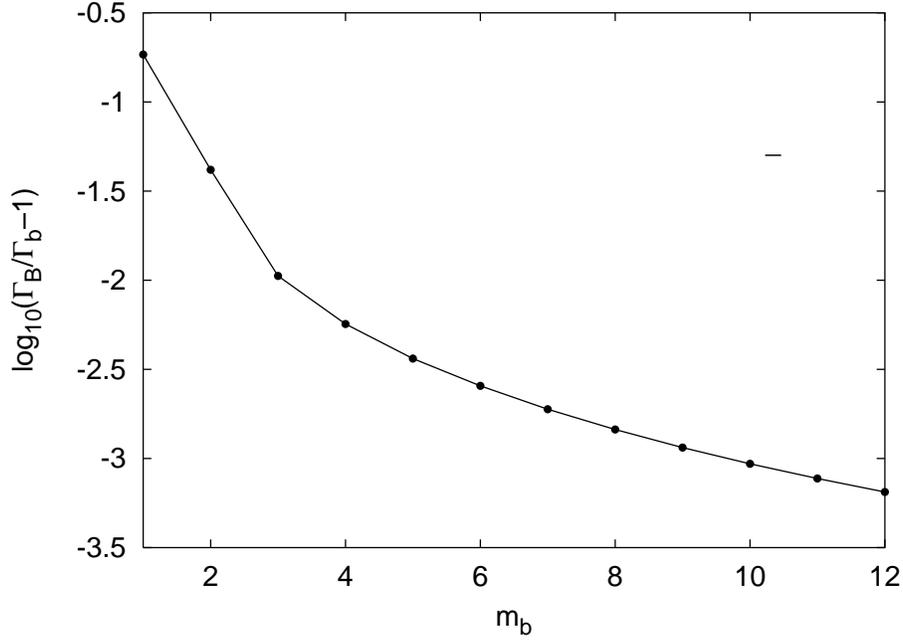}\hfil\hfill

\caption{Same masses as in Fig.~\ref{dual_mc56}, using the Born-term
partonic rate $\Gamma_b$ instead of $\Gamma_{\rm OPE}$.  Note that the
deviation is much larger.}

\label{part_mc56}

\end{centering}
\end{figure}

	The extraordinary agreement between $\Gamma_B$ and
$\Gamma_{\rm OPE}$ may be underscored by instead plotting
(Fig.~\ref{part_mc56}, final column of Table~\ref{dualmc56}) the
difference between $\Gamma_B$ and the Born-term partonic rate
$\Gamma_b$ given in Eq.~(\ref{Born}).  From an algebraic point of
view, $\Gamma_B$ and $\Gamma_{\rm OPE}$ differ generically at
$O(1/M^5)$, while $\Gamma_B$ and $\Gamma_b$ begin to differ already at
$O(1/M^2)$.

	Thus, we find local duality between the actual semileptonic
decay width and its OPE expansion to be very well satisfied in all
cases.

	Before concluding this section, let us briefly address duality
in the differential distribution $\Gamma^{-1}\, {\rm d}\Gamma /{\rm
d}E$.  In the heavy quark limit the shape of the final-state hadronic
mass distribution follows the heavy quark distribution function in the
decaying meson; for the $b \!\to \! u$ decays under consideration,
this is the light-cone distribution function $F(x)$.  In decays with
$q^2\!=\!0$ the recoil energy of the lepton pair $E$ is directly
related to the final state mass $M_h$:
\beq
E = \frac{M_B^2\!-\!M_h^2}{2M_B}\;.
\label{150}
\eeq
Since $q^2\!=\!0$, these decays are analogous to $b \to s \gamma$ in the
Standard Model.  In the large-$m_b$ limit one has
\beq
\frac{1}{\Gamma}\, \frac{{\rm d}\Gamma}{{\rm d}E} \;= \;
\frac{2}{\La}\,
F \left( \frac{2E\!-\!m_b}{\La}\right)\;.
\label{152}
\eeq
At finite $m_b$ in a theory with narrow resonances the actual
distribution is given by the comb of $\delta$-functions with spacing
in the argument of Eq.~(\ref{152}) of order $\La/m_b$.  In order to
obtain a continuous result, we adopt the simple ansatz of averaging
over the peaks.  Using Eq.~(\ref{150}) to define the energy $E_n$ of
the $n$th state $M_h$, we integrate the $\delta$-function for the
$n$th state evenly over the energy range $(E_n \!+\! E_{n+1})/2$ to
$(E_n \!+\! E_{n-1})/2$, i.e., the midpoints between energy
eigenvalues.  Letting $N$ be the maximum number of kinematically
allowed $M_h$ values, we establish the endpoint bins by defining
$E_{-1} \!=\! E_{\rm max} \!=\! M_B/2$ and $E_{N+1} \!=\! E_{\rm min}
\!=\! 0$.

\begin{figure}
  \begin{centering}
  \def\epsfsize#1#2{1.40#2}
  \hfil\epsfbox{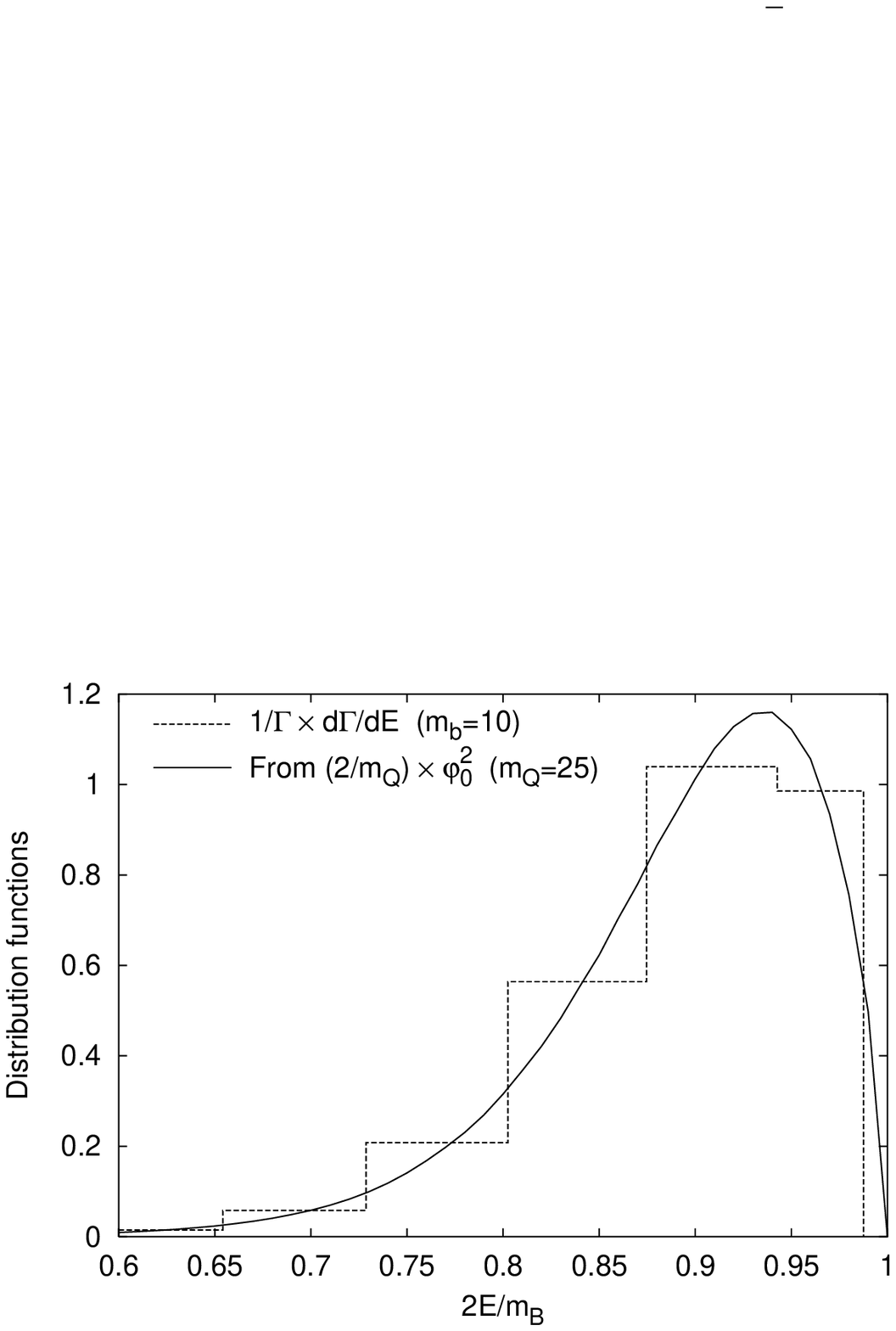}\hfil\hfill

\caption{Exact differential width $\Gamma^{-1}\,{\rm d}\Gamma/{\rm
d}E$, averaged as described in the text, compared to the continuous
parton distribution computed via Eq.~(\ref{diffeq}) with $m_Q \!=\! 25
\beta$, at $m_b \!=\! 10\beta$ and $m\!=\!0.56 \beta$.}

\label{dGdE_25v10}

\end{centering}
\end{figure}

	We find that our numerical computations yield a
distribution resembling the light-cone distribution function
$\varphi^2$; specifically,
\beq
F(y)\;=\;\La\Psi^2 \left((1\!-\!y)\La \right) \;=\;\lim_{
m_Q\to\infty} \frac{\La}{m_Q} \varphi^2
\left(1\!-\!(1\!-\!y)\frac{\La}{m_Q} \right) \;.
\label{154}
\eeq
Recalling that $M_B \!=\! m_b \!+\! \La \!+\! O(1/m_b)$ and combining
Eqs.~(\ref{152}) and (\ref{154}) yields
\begin{equation} \label{diffeq}
\frac{1}{\Gamma} \frac{{\rm d}\Gamma}{{\rm d}E} \approx \lim_{m_Q \to
\infty} \frac{2}{m_Q} \varphi^2 \left( 1 \!-\!
\frac{M_B \!-\! 2E}{m_Q}
\right) .
\end{equation}
The two sides of this expression are plotted in Fig.~\ref{dGdE_25v10},
using $m_Q \!=\! 25\beta$ to represent the limit $m_Q \!\to\! \infty$,
while the actual distribution is considered at $m_b \!=\! 10\beta$.
The agreement is quite remarkable.  The continuous distribution
appears to pass approximately through the midpoint of each bin; owing
to the near-equal spacing of 't~Hooft model eigenvalues in $M_n^2$,
Eq.~(\ref{150}) shows that these bin midpoints are very close to the
values $E_n$ themselves.

\begin{figure}
  \begin{centering}
  \def\epsfsize#1#2{1.38#2}
  \hfil\epsfbox{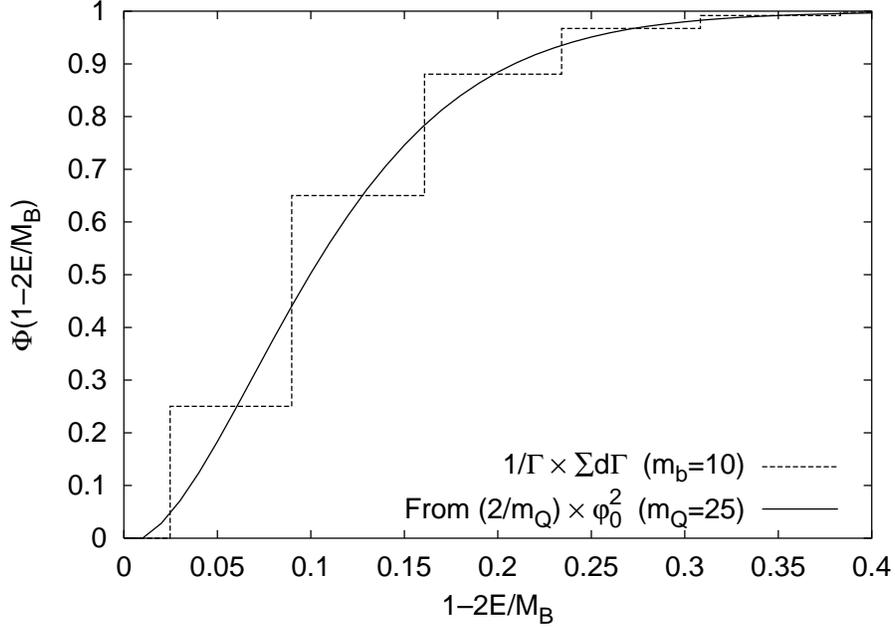}\hfil\hfill

\caption{The partially-integrated (in $E$) differential distribution
$\Phi(1\!-\!2E/M_B)$ and the corresponding smeared exact result
$\Gamma^{-1}\, \sum {\rm d}\Gamma$ as defined in the text, for the
same inputs as in Fig.~\ref{dGdE_25v10}.}

\label{inc10}

\end{centering}
\end{figure}

	It is also interesting to consider integration over a range of
$E$.  In particular, define $\Phi(1\!-\!2E/M_B)$ as the cumulative
fractional width from maximum energy $M_B/2$ down to the given $E$;
then $\Phi(0) \!=\! 0$ and $\Phi(1) \!=\! 1$.  While the exact result
for $\sum \Delta \Gamma/\Gamma$ amounts to an integration of the
$\delta$-function differential widths renormalized so that the
cumulative result approaches unity, the integral of the continuous
distribution gives
\begin{equation} \label{inceq}
\Phi(y) = \lim_{m_Q \to \infty} \:\frac{2}{m_Q} \,\int_0^y {\rm d}z \,
\varphi^2 \left( 1\!-\!\frac{M_{H_Q}}{m_Q} z \right).
\end{equation}
These two curves are presented in Fig.~\ref{inc10}.  Two features
particularly stand out in this plot.  First, even for $m_b$ as large
as $10 \beta \!\approx\! 4\GeV$, the overwhelming part of the decay
probability falls into the transitions to at most four lowest states.
Second, the continuous curve seems to provide a nearly optimal
description possible for the step-like exact distribution.  The
point-to-point deviation for all plotted values with $1\!-\!2E/M_B
\!>\! 0.04$ does not exceed half of the contribution of the nearest
threshold.

\section{Duality in the Vacuum Current Correlator}
\label{vac}

	In this section we briefly illustrate the onset of duality for
the absorptive part of the vector current correlator with light
quarks, of the type that determines the normalized cross section
$R(e^+e^- \to\mbox{hadrons})$ as a function of energy.  In the context
of the heavy quark decays this is relevant in nonleptonic decay widths
in two kinds of processes: in spectator-independent decays, where
$R(q^2)$ determines the weight with which the semileptonic width at
given $q^2$ must be integrated over $q^2$ (see Ref.~\cite{D2WA}), and
in the effects of WA decays.

	In either case, at $N_c \!\to\! \infty$ the cross section
appears as a comb-like collection of $\delta$-functions:
\beq
R(q^2)\;=\; \sum_n c_n^2 \, \delta(q^2\!-\!M_n^2)\;;
\qquad\;
c_n= \int_0^1 {\rm d}x \, \varphi_n(x)\;.
\label{200}
\eeq
The above expression for the residues refers to the case where a
vector current is considered.  We suppress here the factor of
$\sqrt{N_c/\pi}$ relating $c_n$ to $f_n$ [Eq.~(\ref{dec})].  We also
assume in what follows that the light quark masses are equal,
$m_u\!=\!m_d$, and are $O(\beta)$ or less, in order to reach asymptotic
$q^2$ more quickly.

	In the extreme situation of infinitely narrow resonances one
cannot, of course, discuss a point-to-point equality of the cross
section $R(q^2)$ with its OPE in the form of $1/q^2$ expansion.  A
meaningful comparison is possible if each resonant peak is somehow
averaged over an interval no smaller than the distance between
adjacent peaks, the latter being approximately given by $\Delta q^2
\simeq \pi^2\beta^2$ \cite{thooft}.  It is worth recalling that
$R(q^2)$ is proportional to $m^2/q^4$, so one must consider
nonvanishing masses for the vector current, and address the OPE terms
formally suppressed by $m^2/q^2$.

	This question was first addressed in the context of
nonleptonic decays in Ref.~\cite{GL2} using the numerical approach.
Duality for the average cross section in the same manner as above,
i.e., using sum rules derived from the 't~Hooft equation and
analytically matching terms in the $1/m_Q$ expansion, was obtained in
Ref.~\cite{D2WA}.  Yet establishing the asymptotics per se cannot tell
us beforehand how early one can expect the onset of duality.  Here we
study this question numerically, in the domain of intermediate $q^2$.

\begin{figure}
  \begin{centering}
  \def\epsfsize#1#2{1.38#2}
  \hfil\epsfbox{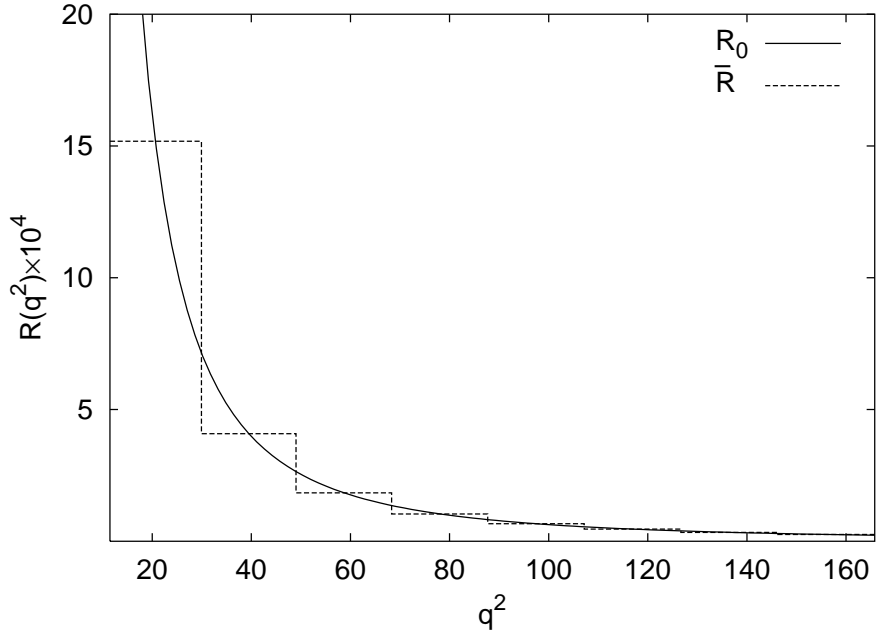}\hfil\hfill

\caption{Saturation of the leading term in the OPE of the vacuum
polarization function $R_0$ by exclusive channel $\delta$-function
contributions, smeared as described in the text ($\overline R$).}

\label{vacfig}

\end{centering}
\end{figure}

	The concrete amount of the deviation between $R(q^2)$ and
$R^{\rm OPE}(q^2)$ in the case of direct resonances may depend in an
essential way on the chosen smearing procedure.  Interested in the
qualitative features only, we choose a rather simplified, crude
method: We spread the integral of $R(q^2)$ evenly over the interval
between the successive resonances.  More precisely, we put
\beq
\bar R(q^2)\;=\; \frac{1}{M_{2n+1}^2\!-\!M_{2n-1}^2}
\int_{M_{2n-1}^2}^{M_{2n+1}^2} {\rm d}q^2 R(q^2)\;=\;
\frac{c_{2n}^2}{M_{2n+1}^2\!-\!M_{2n-1}^2} \; ,
\label{210}
\eeq
for $M_{2n-1}^2 \!<\! q^2 \! <\!  M_{2n+1}^2$, with $M_{-1}^2 \!=\!
4m^2$, the partonic pair production threshold.  Here we use the fact
that $c_n$ vanish for odd $n$ when $m_u \!=\! m_d \equiv m$.  This
smearing is very similar to that described for the differential width
in the last section, except that averaging is performed in $q^2$
rather than $E$.  The free quark loop $R(q^2)$, which is of course the
leading term of the OPE, is given by
\beq
R_0(q^2)\;=\;
\frac{2m^2}{q^4} \frac{1}{\sqrt{1\!-\!4m^2/q^2}}\;.
\label{212}
\eeq
Table~\ref{vaccor} and Fig.~\ref{vacfig} show the results for our
reference case $m \!=\! 0.56 \beta$.  The agreement of the average
hadronic cross section with the parton-computed probability again
turns out to be very good.  Apparently, this can be related to two
facts: the heavy suppression of power corrections to $R(q^2)$ in the
OPE (see Eqs.~(34)--(35) in \cite{D2WA}), and an early onset of the
asymptotics in the spectrum,
\begin{equation}
M_{n+1}^2\!-\!M_n^2 \:\simeq\: \pi^2\beta^2\; ,
\end{equation}
which even at $n\!=\!0$ is satisfied to about $15\%$.

\section{Discussion and Summary}
\label{concl}

	The main motivation behind the present study has been to
assess the magnitude of local duality violations in the inclusive
semileptonic decays of beauty particles.  We considered this question
using the 't~Hooft model as a toy theory in which all relevant decay
amplitudes can be evaluated numerically.  The 't~Hooft model, while
retaining certain key features of full $D\!=\!4$ QCD that shape the
spectrum of hadrons (quark confinement, chiral symmetry breaking),
still differs from $D\!=\!4$ in many respects.  Yet using it as a lab
for exploration carries an important advantage---it allows no ``wiggle
room'' for interpretation of the results.  There are no ad hoc
parameters to choose or adjust, and as soon as the underlying weak
decay Lagrangian is fixed, the numerical results are unambiguous and
must be accepted at face value.  This positively distinguishes this
approach from various models where often the conclusions, even
qualitatively, depend on the arbitrary choice of parameters according
to one's preferences.  The question of a particular model being
compatible with the general dynamical properties of QCD underlying the
OPE approach, often quite problematic in simplified quark models, does
not arise for the 't~Hooft model.

	Although the simplest illustration of the asymptotic nature of
the decay width $1/m_Q$ expansion and related violations of local
duality \cite{shifdual} follows just from the existence of hadronic
thresholds (see, e.g., \cite{D2WA}), violation of local duality is a
more universal phenomenon that is {\em not\/} directly related to
existence of hadronic resonances nor even confinement itself.  This
has been illustrated in Ref.~\cite{inst} by the example of soft
instanton effects that do not lead, at least at small density, to
quark confinement---but do indeed generate computable oscillating
duality-violating contributions to the total decay rates.

	Nevertheless, there is a widespread opinion that decays with
manifest resonance structure in the final state are most difficult
for---if compatible at all with---the standard OPE.  Even the
possibility that the OPE does not fully apply in the case of ``hard''
confinement has been occasionally voiced in the literature.  The
analytic studies performed in Refs.~\cite{D2,D2WA,D2PI}, which
explicitly demonstrate in the 't~Hooft model the applicability of the
OPE to the total widths, should help to allay such conceptual
concerns.  Nevertheless, the intuition remains that resonance
dominance is not ``favorable'' for the OPE, and problems might show
up, for instance, through a delayed numerical onset of duality, in
that the approximate equality of the OPE predictions and the actual
decay widths may set in only after a significant number of thresholds
has been passed.  To address such issues, the 't-Hooft model seems to
represent the most certain testing ground for local duality in the
domain of decays of moderately heavy quarks.

	Contrary to naive expectations, we found surprisingly accurate
duality between the (truncated) OPE series for $\Gamma_{\rm sl}$ and
the actual decay widths.  The deviations are suppressed to a very high
degree almost immediately after the threshold for the first excited
final state hadron is passed.  No suspected delay in the onset of
duality was found.

	The key property that governs the onset of the $1/m_Q$
expansion for the semileptonic widths is the pattern of saturation of
the heavy quark sum rules.  We examined a particular class, the SV sum
rules in the heavy quark limit, that has the most transparent quantum
mechanical meaning.  We found them saturated to an amazing degree by
the very first excitation.  The contribution of the remaining, higher
states to the slope of the IW function, $\La$, and $\mu_\pi^2$ does
not exceed a few percent.  Even in the Darwin operator sum rule, the
first excitation accounts for $90\%$ of the whole expectation value,
despite the fast-growing weight, $(\epsilon_k\!-\!\epsilon_0)^3$ of
higher-order contributions.  This peculiarity underlies the early
onset of duality for the case when initial- and final-state quarks are
both heavy.

	Some of the duality-violating features observed in these
studies have natural explanations.  At fixed energy release
$m_Q\!-m_q$ the magnitude of the deviations is smaller if $m_Q$,
$m_q$ are both large (as in $b \!\to\! c$) than if they are both
small.  This is expected, since in the former case the heavy quark
symmetry for the elastic amplitude additionally enforces approximate
duality even when no expansion in large energy release can be applied.

	It is interesting, however, that at fixed $m_b$ the duality
violation decreases rapidly as $m_c$ decreases, in full accord with
the OPE where the higher order terms are generally suppressed by
powers of $1/(m_b\!-\!m_c)$.  This is clearly a {\em dynamical\/}
feature that goes beyond heavy quark symmetry per se, the quality of
which deteriorates as $m_c$ decreases.

	It is also instructive to note that including the calculated
power-suppressed OPE terms significantly reduces the difference
between the actual decay width and its purely partonic evaluation.
Moreover, the seeds of oscillations inherent to duality violation (as
functions of quark masses), can be seen.  Since we adopted the
truncated OPE expansion to mirror the existing implementation of the
OPE in QCD, the deviations do not average to zero but rather oscillate
around the (rapidly dissipating) contributions attributed to discarded
higher-order terms.

	The numerical effects of duality violation we study turn out
to be typically quite small.  Partially this can be attributed to
moderate size of the corresponding expectation values multiplying
$1/m_Q^k$ corrections in the OPE.  Yet certainly not all power
corrections in heavy quarks are suppressed in the model.  It is well
known from ordinary quantum mechanics that masses (eigenvalues)
typically are much more robust against perturbations than
wavefunctions themselves (or transition amplitudes).  We observe a
similar pattern in the 't~Hooft model.  For example, $1/m_Q$
corrections to the meson decay constants turn out very significant
even at the scale of the $b$ quark mass.  Apparently, the inclusive
decay rates fall into the class of ``robust'' observables, although,
as explained above, this was difficult to anticipate beforehand.

	We note here another ``fragile'' observable, the light-cone
heavy quark distribution function, which can be measured in decays of
the type $b\to s \gamma$.  In $D\!=\!2$ the scaled spread
$m_Q^2\left(\aver{x^2}\!-\!\aver{x}^2\right)$ of the $x$ distribution
approaches $\mu_\pi^2$ at large $m_Q$.  Yet, as seen in
Fig.~\ref{Mx2}, even at the $b$ quark mass one would obtain from this
distribution only about $60\%$ of the actual value of $\mu_\pi^2$, due
to significant $1/m_Q$ corrections.  This caveat may be important for
existing analyses of the decay distributions in $B$ decays, where such
effects routinely are not included.

	We also briefly addressed the inclusive differential decay
distributions in the analogues of $b\to u\,\ell \bar\nu$ or $b\to s
\gamma$ decays.  Generally, we find good agreement (at the scale
corresponding to the physical $b$ mass) with the parton-based
prediction incorporating effects of the ``Fermi motion,'' and in
particular for the partially integrated probability
\beq \Phi(x) \,=\, \frac{1}{\Gamma_{\rm sl}} \,\int_0^{xM_B^2}
{\rm d}M_h^2 \: \frac{{\rm d}\Gamma_{\rm sl}}{{\rm d}M_h^2}\;.
\label{C4}
\eeq
This distribution, following Refs.~\cite{dist,mx}, is examined in real
$B$ decays in the quest for $|V_{ub}|$ \cite{yelbook}.  However, the
point-to-point deviations are clearly still significant, for the
decays to only the 4 or 5 lowest final states saturate the
overwhelming fraction of the total decay probability.  It is quite
conceivable, though, that such deviations are less pronounced in
actual QCD owing to the significant resonance widths and to a richer
resonance structure.

	The vacuum current correlator also turns out to be especially
robust; even neglecting all OPE corrections except the leading
partonic contribution leads to excellent agreement with the hadronic
result.

	Turning to the direct phenomenological conclusions that can be
inferred from our studies, we see that, to the extent our findings can
be transferred to real QCD, violation of local duality in the total
semileptonic widths of $B$ mesons is not an issue.  The scale of
duality violation lies far below the phenomenologically accessible
limits, and cannot affect the credibility of $|V_{cb}|$ or $|V_{ub}|$
extractions.

	In reality there are, of course, essential conceptual
differences between the two theories, including those aspects that are
expected to be essential for local duality (for a discussion, see
Ref.~\cite{D2WA}).  Although many seem to optimistically suggest that
duality violation is more pronounced in the 't~Hooft model than for
actual heavy flavor hadrons, some differences may still work in the
opposite direction.  In $D\!=\!2$ there are no dynamical gluons, nor a
chromomagnetic field that in $D\!=\!4$ provides a significant scale of
nonperturbative effects in heavy flavor hadrons.  Likewise, there is
no spin in $D\!=\!2$, and no corresponding $P$-wave excitations of the
light degrees of freedom (the so-called $j\!=\!3/2$ states), which
seem to play an important role in $D\!=\!4$.

	Two-dimensional QCD neither has long perturbative ``tails'' of
actual strong interactions suppressed weakly (by only powers of
$\log$s of the energy scale).  In $D\!=\!2$ the perturbative
corrections are generally power-suppressed, as follows from the
dimension of the gauge coupling.  As discussed in Ref.~\cite{D2WA}, it
is conceivable that the characteristic mass scale for freezing out the
transverse gluonic degrees of freedom is higher than in the
``valence'' quark channels.  This would imply a possibly higher scale
for onset of duality in $\alpha_s/\pi$ corrections to various
observables.

	Regardless of these differences, we conclude that presence of
resonance structure per se is not an obstacle for fine local
quark-hadron duality tested in the context of the OPE.  As we see in
the 't~Hooft model, resonances themselves do not seem to demand a
larger duality interval.  As soon as the mass scale of the states
saturating the sum rules in a particular channel (quark or hybrid) has
been passed, the decay width can be well approximated numerically by
the expansion stemming from the OPE.

	The ground states of heavy mesons in the 't~Hooft model
exhibit relatively small expectation values of nonperturbative
operators ($\mu_\pi^2$, $\rho_D^3$, but not $\La$) compared to real
QCD, if our identification $\beta\simeq 400\MeV$ is adopted.  This may
be regarded as a reason for small duality violation for $\Gamma_{\rm
sl}$ in the model.  However, even if we scale $\beta$ up to
$700$--$800 \MeV$ to make up for smallness of the nonperturbative OPE
effects, the duality violation is still very small, and superficially
rather insignificant even in charm.

	We note, however, that the specific choice Eq.~(\ref{120}) of
the weak interaction effectively requires decays to occur only at
$q^2\!=\!0$, and therefore the effects of four-fermion operators of
the type $(\bar Q \Gamma q)(\bar q \Gamma Q)$ are totally absent, at
least in the lowest orders of perturbation theory (cf.\
Ref.~\cite{D2}, Sec.~III.B.3).  As was suggested in Ref.~\cite{vub},
it is conceivable that the apparent excess in $\Gamma_{\rm sl}(D)$ is
simply related to a noticeable magnitude of the non-valence
(nonfactorizable) expectation values $\matel{D}{(\bar{c}\Gamma
s)\,(\bar{s}\Gamma c)}{D}$.  If this conjecture is true, similar
effects in $\Gamma_{\rm sl}(B\to X_u\ell\nu)$ are still suppressed but
possibly detectable in future precision experiments.  In the context
of the present study, it suffices to say that this would be a
legitimate OPE effect rather than a manifestation of a significant
local duality violation in the strict sense.
\vspace*{.3cm}\\
{\bf Acknowledgments:}~
R.L.\ thanks the Department of Energy for support under Contract No.\
DE-AC05-84ER40150; N.U.\ acknowledges the support of the NSF under
grant number PHY96-05080, by NATO under the reference PST.CLG 974745,
and by RFFI under grant No.\ 99-02-18355.  We are grateful to N.~Isgur
for inspiring interest and discussions, and to M.~Burkardt for
invaluable insights.  N.U.\ also thanks I.~Bigi, M.~Shifman and
A.~Vainshtein for encouraging interest and collaboration on related
issues, and A.\ Zhitnitsky for useful comments.  N.U.\ enjoyed the
hospitality of Physics Department of the Technion and the support of
the Lady Davis grant during completion of this paper.

\appendix

\section{The BSW Improvement of the Multhopp Technique}
\label{newmult}

        The Brower-Spence-Weis \cite{BSW} (BSW) improvement of the
Multhopp technique avoids the need for evaluating the wavefunction at
a discrete set of points called ``Multhopp angles,'' thus improving
the behavior of the solutions in the endpoint regions, as described in
Sec.~\ref{tHm}.  Here we exhibit the expressions used by BSW,
correcting along the way some minor typographical errors in their
work.

        Starting with the 't~Hooft equation (\ref{tHe}) with bare
quark masses $m_1$ and $m_2$, one converts the kinematic variables
$x,y$ to angular variables:
\begin{equation}
x = \frac{1\!+\!\cos {\theta}}{2}, \qquad
y = \frac{1\!+\!\cos {\theta^\prime}}{2} \;,
\end{equation}
in terms of which the 't~Hooft equation reads
\begin{eqnarray}
\frac{M_p^2}{2} \, \varphi_p (\theta) & = &
\left[ \frac{m_1^2}{1\!+\!\cos{\theta}} +
\frac{m_2^2}{1\!-\!\cos{\theta}} \right]
\varphi_p (\theta) \nonumber
\\ & & +
\int_0^\pi
{\rm d}\theta^\prime \, \varphi_p (\theta^\prime)
\, {\rm P} \frac{1}{(\cos {\theta} \!-\! \cos {\theta^\prime})^2} .
\end{eqnarray}
Expanding
\begin{equation} \label{exp}
\varphi_p (\theta) \;=\; \sum_{n=1}^{\infty} \,a_n^{(p)}
\sin {n\theta} ,
\end{equation}
and using the continuous inversion identity (contrast with Eq.~(A7) of
Ref.~\cite{GL1})
\begin{equation}
\int_0^\pi {\rm d}\theta \: \sin {m\theta}\, \sin {n\theta}
\;=\;
\frac{\pi}{2} \left( \delta_{mn} \!-\! \delta_{m,-n} \right) ,
\end{equation}
one obtains the infinite-dimensional eigenvector system
\begin{equation} \label{MulBSW}
M_p^2 a_n^{(p)} \;=\; \left( H_0 + V \right)_{nm} a_m^{(p)} \;,
\end{equation}
where
\begin{eqnarray}
\left( H_0 \right)_{nm} & = & + \frac{4}{\pi} \int_0^\pi {\rm d}\theta
\, \left[ \frac{m_1^2}{1\!+\!\cos{\theta}} +
\frac{m_2^2}{1\!-\!\cos{\theta}}
\right]\, \sin {n\theta} \,\sin {m\theta} , \\
V_{nm} & = & -\frac{4}{\pi} \beta^2 \int_0^\pi {\rm d}\theta
\, \sin {n\theta}\, \int_0^\pi
{\rm d}\theta^\prime \,\sin{\theta^\prime}
\sin {m\theta^\prime} \, {\rm P}
\frac{1}{(\cos {\theta} \!-\! \cos{\theta^\prime})^2} \;.
\label{vmn1}
\end{eqnarray}
Both of these integrals can be evaluated, with the result
\begin{eqnarray}
\left( H_0 \right)_{nm} & = & 4 \, \min (n,m) \left[ (-1)^{m+n} m_1^2
+ m_2^2 \right] , \\
V_{nm} & = & V_{n-1,m-1} \left( \frac{m}{m\!-\!1} \right) +
\frac{8m}{n\!+\!m\!-\!1} \left[ \frac{1+(-1)^{n+m}}{2} \right] ,
\end{eqnarray}
where $V_{n0} \!=\! V_{0m} \!=\! 0$ for $m,n \ge 0$.  This recursive
form for $V_{nm}$ is most convenient for numerical calculations;
however, one may also write the closed-form solution,
\begin{equation} \label{vmn2}
V_{nm} = 4m \left[ \frac{1+(-1)^{n+m}}{2} \right]
\left[ \psi \left(
\frac{1\!-\!n\!-\!m^{\vphantom{2}}}{2} \right) -
\psi \left( \frac{1\!-\!
|n\!-\!m|}{2} \right) \right] .
\end{equation}
Note that the ``potential'' $V$ in Eq.~(\ref{vmn2}) is real but not
symmetric, owing to the extra $\sin \theta^\prime$ in
Eq.~(\ref{vmn1}); therefore, the ``Hamiltonian'' $H_0\!+\!V$ is not
Hermitian, and the eigenvectors $a^{(p)}$ are not orthogonal.  This is
a direct result of converting the exact wavefunctions, which are
eigenfunctions of a Hermitian Hamiltonian when written in terms of the
variable $x$ (and therefore orthogonal in $x$), into orthogonal
functions of the variable $\theta$.  This transformation is nonunitary
because the number of modes used is not infinite; therefore, the
overlap of different eigenvector solutions should be small when a
large number of modes are used.  Indeed, this turns out to be
empirically true; nevertheless, we take the further step of
orthogonalizing the numerical eigenvector solutions recursively by
means of the standard Gram-Schmidt procedure, i.e.,
\begin{equation}
\left| \varphi^{(p)}_{\rm orth} \right> = \frac{ \left| \varphi^{(p)}
\right> - \sum_{j=0}^{p-1} \left| \varphi^{(j)}_{\rm orth} \right>
\left< \varphi^{(j)}_{\rm orth} \right| \left. \varphi^{(p)} \right> }
{\sqrt{\left< \varphi^{(p)}_{\vphantom{x}} \right|
\left. \varphi^{(p)}_{\vphantom{x}} \right> - \sum_{j=0}^{p-1} \left<
\varphi^{(j)}_{\rm orth} \right| \left. \varphi^{(p)}_{\vphantom{x}}
\right> }} .
\end{equation}
For $N=500$ modes, this typically changes expectation values by one
part in $10^5$.

        The expressions for these overlaps and other matrix elements
in terms of the mode coefficients $a_n$ are presented in
Appendix~\ref{ovlap}.

\section{Matrix Elements}
\label{ovlap}

        A number of useful overlaps and other integrals are
straightforward to evaluate in terms of the mode coefficients, using
the expressions (\ref{exp}).  Solving them amounts to evaluating a
number of trigonometric integrals.  Such expressions are especially
convenient since they permit a number of integrations that introduce
no numerical uncertainties (except due to machine precision) beyond
those of solving the original Multhopp-BSW eigenvector equation
(\ref{MulBSW}).

        In particular, denote the $p$th eigenstate wavefunction
presented in Eq.~(\ref{exp}) by $\varphi^{(a)}_p$ and that for some
other set of masses in the $q$th eigenstate by $\varphi^{(b)}_q$; the
latter wavefunction then has an expansion like (\ref{exp}) with mode
coefficients $b^{(q)}_n$.  Truncating after $N$ modes, one then finds
\begin{eqnarray}
\left< \varphi_p^{(a)} \right| \left. \, \varphi_q^{(b)} \right> & = &
\int_0^1 {\rm d}x \, \varphi_p^{(a)} (x) \varphi_q^{(b)} (x) \nonumber
\\ & = & -2 \sum_{m=1}^N m a_m^{(p)} \sum_{n=1}^N n b_n^{(q)} \left[
\frac{1+(-1)^{m+n}}{2} \right]
\frac{1}{\left[ 1 \!-\! (m\!-\!n)^2 \right]
\left[ 1 \!-\! (m\!+\!n)^2 \right]} . \nonumber \\ & &
\end{eqnarray}
Indeed, the normalization integral $\int_0^1 {\rm d}x \, \varphi(x)^2
\!=\! 1$ is just the case $a\!=\!b$ and $p\!=\!q$, in agreement with
Eq.~(A9) of Ref.~\cite{GL1}.

        Other useful expectation values include
\begin{eqnarray}
\left< x \!-\! \frac 1 2 \right >_p & = &
\int_0^1 {\rm d}x \, \left( x\!-\!
\frac 1 2 \right) \left[ \varphi^{(a)}_p (x) \right]^2 \nonumber \\ &
= & -\sum_{m=1}^N m a_m^{(p)} \sum_{n=1}^N n a_n^{(p)}
\left[ \frac{1
\!-\! (-1)^{m+n}}{2} \right]
\frac{1}{\left[ 4 \!-\! (m\!-\!n)^2 \right] \left[ 4
\!-\! (m\!+\!n)^2 \right]} , \nonumber \\ & &
\end{eqnarray}
\begin{eqnarray}
\lefteqn{\left< \left( x \!-\! \frac 1 2 \right)^2 \right>_p =
\int_0^1
{\rm d}x \,
\left( x\!-\! \frac 1 2 \right)^2 \left[ \varphi^{(a)}_p (x)
\right]^2 \nonumber} \\
& = & -\frac 1 2 \sum_{m=1}^N m a_m^{(p)}
\sum_{n=1}^N n a_n^{(p)}
\left[ \frac{1 \!+\! (-1)^{m+n}}{2} \right]
\left[ 21 \!-\! 6(m^2\!+\!n^2) + (m^2\!-\!n^2)^2 \right]
\nonumber \\ & & \times
\left[ \left( 1 \!-\! (m\!-\!n)^2 \right) \left( 1 \!-\!
(m\!+\!n)^2 \right)
\left( 9 \!-\!
(m\!-\!n)^2 \right) \left( 9 \!-\! (m\!+\!n)^2 \right) \right]^{-1} .
\end{eqnarray}
Note that the spread of the wavefunction may be computed about any
convenient point in $x$, viz.,
\begin{equation}
\left< \left( ax\!+\!b \right)^2 \right> -
\left< \left( ax\!+\!b
\right)^{\vphantom{2}} \right>^2 = a^2 \left( \left< x^2 \right>
\!-\!
\left< x^{\vphantom{2}} \right>^2 \right) ,
\end{equation}
so that the additive constants of $-1/2$ above are irrelevant.  Also,
\begin{equation}
\left< \frac 1 x \right>_p \,=\, \int_0^1 {\rm d}x \, \frac 1 x \left[
\varphi_p^{(a)} (x) \right]^2 \:=\: \sum_{m=1}^N a_m^{(p)}
\,\sum_{n=1}^N a_n^{(p)} \, I_{mn} ,
\end{equation}
where
\begin{eqnarray}
I_{mn} & = & 2 \sum_{j=|m-n|/2}^{(m+n)/2-1} \frac{1}{2j\!+\!1}
\nonumber
\\ & = &
\psi \left( \frac{m\!+\!n^{\vphantom{2}}\!+\!1}{2} \right) - \psi
\left ( \frac{|m\!-\!n|\!+\!1}{2} \right) , \ \ m\!-\!n \ {\rm even};
\nonumber \\
I_{mn} & = & \frac{1}{|m\!-\!n|} - \frac{1}{m\!+\!n} - 2
\sum_{j=(|m-n|-1)/2}^{(m+n-1)/2 -1} \frac{1}{2j\!+\!1}
\nonumber \\ & = &
\frac{1}{|m\!-\!n|} - \frac{1}{m\!+\!n} +
\psi \left( \frac{|m\!-\!n|}{2}
\right) - \psi \left( \frac{m\!+\!n^{\vphantom{\prime}}}{2} \right) , \
\ m\!-\!n \ {\rm odd}.
\label{Imn}
\end{eqnarray}
One also finds
\begin{equation}
\left< \frac{1}{1\!-\!x} \right>_p \:=\: \int_0^1 {\rm d}x \,
\frac{1}{1\!-\!x}
\left[ \varphi_p^{(a)} (x) \right]^2 \,=\, \sum_{m=1}^N \,a_m^{(p)}
\,\sum_{n=1}^N \,a_n^{(p)} \, J_{mn} \;,
\end{equation}
where, using the notation of Eq.~(\ref{Imn}), one finds $J_{mn} \!=\!
+I_{mn}$ for $m\!-\!n$ even, and $J_{mn} \!=\! -I_{mn}$ for $m\!-\!n$
odd.

        Finally, the decay constant of the $p$th excitation [cf.\
Eqs.~(\ref{112})--(\ref{cn})] is given by
\begin{equation} \label{dec}
f_p^{(a)} = \sqrt{\frac{N_c}{\pi}} \int_0^1 {\rm d}x \,
\varphi_p^{(a)} (x) = \sqrt{\frac{N_c}{\pi}} \, c_p =
\sqrt{\frac{N_c}{\pi}} \times \frac{\pi}{4} \, a_1^{(p)} \:.
\end{equation}

\section{Additional Relations Used in the Analysis}
\label{calc}

	The numerical calculation of large-$m_Q$ matrix elements with
acceptable accuracy relies on achieving a balance between competing
effects.

	On one hand, Multhopp solutions to the 't~Hooft equation with
$m_Q \gg \beta$ tend to suffer degraded numerical accuracy since they
are highly concentrated into the small kinematic region $1\!-\!x \ll
1$.  As discussed in Sec.~\ref{tHm}, the endpoint regions $x \!\approx
\!0$ and $1$ are where the Multhopp solutions---or more precisely,
their derivatives---tend to break down.  This effect is compounded
when $m \ll \beta$, since lighter quark masses force sharper endpoint
behavior in the wavefunction.  Although the BSW solution ameliorates
this behavior, as $m_Q$ is increased one eventually faces the problem
of attempting to represent a function with only a very small region of
support in $x$ by a finite number of modes with support over the full
range $x \in [0,1]$.  In practice, we gauge the errors committed
through such ``lattice spacing'' effects by computing a given quantity
with $N \!=\! 500$ and noting the amount by which its value shifts if
one uses instead $N \!=\! 100$, and as expected, such errors become
substantial (as much as a few percent) by the time one reaches $m_Q
\!>\! 25 \beta$ or $m \!<\! 0.4 \beta$.

	On the other hand, although numerical solutions with $m_Q,\, m
\simeq O(\beta)$ have the highest numerical accuracy, they also
have substantial $O(1/m_Q)$, $O(1/m_Q^2)$, etc.\
corrections that are difficult to disentangle.

	We adopt an intermediate strategy of employing certain exact
relations that hold for the 't~Hooft solutions.  To determine the
relevant static expectation values, we solve the finite-$m_Q$ heavy
hadron mass expansion for $\La\,$ [Eq.~(\ref{50})]:
\begin{equation}
M_{H_Q}\!-\!m_Q\,=\, \La +
\frac{\mu_\pi^2\!-\!\beta^2}{2m_Q}  +
\frac{\rho_D^3-\rho_{\pi\pi}^3}{4m_Q^2} \;+\;
O\!\left(\frac{\beta^4}{m_Q^3}\right)
\,.
\label{C50}
\end{equation}
Neglecting the order term and using the relations [Eq.~(\ref{52})]
\begin{equation}
\mu_\pi^2= \frac{\La^2\!-\!m^2+\beta^2}{3}\;,\qquad \rho_D^3=
\frac{\beta^2 F^2}{4}\;,\qquad \rho_{\pi\pi}^3 = \frac{1}{36} \left[
8\La (\La^2\!-\!m^2\!+\!\beta^2) + 3 \beta^2 F^2 \right] ,
\label{C52}
\end{equation}
we thus arrive at an equation cubic in $\La$ that depends on $F^2$.
We solve it at $m_Q\!=\!15\beta$.

	The asymptotic value of the scaled decay constant
$F^{(n)}\!=\!\sqrt{m_Q}\, c_n$ must also be evaluated at a finite
value of $m_Q$, thus including $1/m_Q$-suppressed pieces.  We account
for them explicitly using the expansion \cite{burkur} [the first of
Eqs.~(\ref{58})]
\begin{equation}
\sqrt{m_Q} \, c_n = \left( 1 - \frac{2[2\La^{(n)} \!-\!
m(-1)^{n}]}{3m_Q}
\right) F^{(n)} + O\left(\frac{\beta^{5/2}}{m_Q^2}\right)\;.
\label{C58}
\end{equation}
We likewise solve this equation for $F^{(n)}$ at $m_Q\!=\!15\beta$.

	Turning to the analysis of the SV sum rules
Eqs.~(\ref{oscsum1})--(\ref{oscsum4}) in Sec.~4, we note that their
rapid saturation demands an exceptionally high precision in evaluating
both the oscillation strengths $\tau$ in the r.h.s.\ and the
expectation values in the l.h.s\@.  Reaching such an accuracy through
direct computation seems impossible.  Therefore, we use a number of
identities to get meaningful results.  First, we employ the expression
for $\tau_{nk}$ in terms of $\epsilon_n$, $\epsilon_k$, and the
corresponding decay constants:
\begin{equation}
\tau_{nk} =
- \frac{\beta^2}{2(\epsilon_n\!-\!\epsilon_k)^3} \, F^{(n)} F^{(k)}
\left(\frac{1\!-\! (-1)^{n-k}}{2} \right) \;.
\label{C110}
\end{equation}
Then we make use of the fact that the discussed sum rules, being
completeness sums, are exact when summation includes all excitations
(see Ref.~\cite{burkur}).  Therefore, one has
\bea
\frac{1}{\rho_k^2\!-\!\frac{1}{4}}\;
\left[\rho_k^2\!-\!\frac{1}{4} - \sum_{\ell=1}^{n} \tau_{\ell k}^2 \,
\right] &=&
\frac{1}{\rho_k^2\!-\!\frac{1}{4}}\:
\sum_{\ell=n+1}^\infty  \tau_{\ell k}^2 \; ,
\label{cSV1}
\\
\frac{2}{\La_k}\; \left[
\frac{1}{2} \La_k - \sum_{\ell=1}^{n}
(\epsilon_\ell\!-\!\epsilon_k) \;\tau_{\ell k}^2 \,\right]
&=&
\;\,\frac{2}{\La_k}\;\:
\sum_{\ell=n+1}^\infty  (\epsilon_\ell\!-\!\epsilon_k) \;
\tau_{\ell k}^2 \; , \\
\frac{1}{\left(\mu_\pi^2\right)_k}
\left[\left(\mu_\pi^2\right)_k - \sum_{\ell=1}^{n}
(\epsilon_\ell\!-\!\epsilon_k)^2 \,\tau_{\ell k}^2 \,\right]
&=&
\frac{1}{\left(\mu_\pi^2\right)_k}
\sum_{\ell=n+1}^\infty (\epsilon_\ell\!-\!\epsilon_k)^2 \,
\tau_{\ell k}^2 \; ,
\\
\frac{1}{\left(\rho_D^3\right)_k}
\left[
\left(\rho_D^3\right)_k - \sum_{\ell=1}^{n}
(\epsilon_\ell\!-\!\epsilon_k)^3 \,\tau_{\ell k}^2 \,\right]&=&
\frac{1}{\left(\rho_D^3\right)_k}
\sum_{\ell=n+1}^\infty (\epsilon_\ell\!-\!\epsilon_k)^3\,
\tau_{\ell k}^2 \; .
\label{CSV$}
\eea
The sums on the r.h.s.\ can be accurately evaluated since the higher
contributions fall off in magnitude very quickly.  In practice, we
truncate the sum at $\ell\!=\!20$.

	A similar approach was used to evaluate the duality-violating
difference $\Gamma_B-\Gamma_{\rm OPE}$ as a function of $m_Q$.  We use
the exact relation \cite{D2} [Eqs.~(\ref{124}), (\ref{130}),
(\ref{141})]
\begin{equation}
\Gamma_B= \frac{G^2}{4\pi}\cdot \frac{m_b^2 \!-\!m_c^2}{M_B}
\int _0^1 \frac{{\rm d}x}{x}
\,\varphi_B^2(x) \,- \,
 \frac{G^2}{4\pi}
\sum _{M_n> M_B}
\frac{M_B^2\! -\! M_n^2}{M_B}
\left| \int _0^1 {\rm d}x \, \varphi_n(x)\varphi_B(x) \right|^2  ,
\label{C130}
\end{equation}
and therefore,
\begin{equation}
\frac{\Gamma_B\!-\!\Gamma_{\rm OPE}}{\Gamma_{\rm OPE}}
=
\left(\int _0^1 \frac{{\rm d}x}{x}\,\varphi_B^2(x) \right)^{-1}\,
\sum_n \; \frac{M_n^2\!-\!M_B^2}{m_b^2\!-\!m_c^2}
\,\left| \int _0^1 {\rm d}x \, \varphi_n(x)\varphi_B(x) \right|^2
\theta(M_n\!-\!M_B)\;.
\label{C132}
\end{equation}
The summation runs over all final excited states kinematically {\em
forbidden\/} in the decay.  Once again, the sum converges rapidly and
is dominated by the lowest couple of states.

\clearpage

\begin{table}

\begin{centering}

\begin{tabular}{c|c|c|c|c|c}

$m_Q$ & $M_{H_Q} \!-\! m_Q$ &
$\sqrt{m_Q} |c_n|$ & $m_Q \langle 1\!-\!x \rangle $ &
$m_Q^2 \left( \langle x^2 \rangle \!-\! \langle x \rangle^2 \right)$ &
$\frac{m_Q}{M_{H_Q}} \langle \frac 1 x \rangle \!-\! 1$ \\
\hline \multicolumn{6}{c}{Ground state $(n\!=\!0)$} \\ \hline

0.56 & 1.21918 & 0.7300 & 0.280 & 0.017 & $8.34 \times
  10^{-2}$ \\

 1.0 & 1.24633 & 0.9534 & 0.432 & 0.048 & $4.59 \times
  10^{-2}$ \\

 3.0 & 1.28764 & 1.4210 & 0.791 & 0.211 & $9.04 \times
  10^{-3}$ \\

 5.0 & 1.29904 & 1.6061 & 0.944 & 0.333 & $3.58 \times
  10^{-3}$ \\

 7.0 & 1.30423 & 1.7503 & 1.029 & 0.417 & $1.88 \times
   10^{-3}$ \\

10.0 & 1.30820 & 1.7901 & 1.102 & 0.500 & $9.33 \times
   10^{-4}$ \\

15.0 & 1.31131 & 1.8633 & 1.166 & 0.582 & $4.14 \times
   10^{-4}$ \\

25.0 & 1.31375 & 1.9271 & 1.222 & 0.661 & $1.46 \times
   10^{-4}$ \\

35.0 & 1.31475 & 1.9560 & 1.248 & 0.700 & $7.31 \times
   10^{-5}$ \\

50.0 & 1.31545 & 1.9783 & 1.268 & 0.732 & $3.53 \times
   10^{-5}$ \\
\hline \multicolumn{6}{c}{First excited state $(n\!=\!1)$} \\ \hline

0.56 & 2.82831 & 0.0000 & 0.280 & 0.032 & $-1.30 \times
  10^{-1}$ \\

 1.0 & 2.77888 & 0.0922 & 0.476 & 0.091 & $-9.62 \times
   10^{-2}$ \\

 3.0 & 2.66569 & 0.4429 & 1.094 & 0.457 & $-3.29 \times
   10^{-2}$ \\

 5.0 & 2.61977 & 0.6427 & 1.437 & 0.775 & $-1.60 \times
   10^{-2}$ \\

 7.0 & 2.59522 & 0.7648 & 1.649 & 1.014 & $-9.40 \times
   10^{-3}$ \\

10.0 & 2.57436 & 0.8775 & 1.848 & 1.267 & $-5.14 \times
   10^{-3}$ \\

15.0 & 2.55644 & 0.9812 & 2.033 & 1.529 & $-2.50 \times
   10^{-3}$ \\

25.0 & 2.54088 & 1.0765 & 2.205 & 1.797 & $-9.68 \times
   10^{-4}$ \\

35.0 & 2.53382 & 1.1213 & 2.287 & 1.933 & $-5.10 \times
   10^{-4}$ \\

50.0 & 2.52833 & 1.1566 & 2.351 & 2.046 & $-2.56 \times
   10^{-4}$

\end{tabular}

\caption{Matrix elements as functions of heavy quark mass $m_Q$ and
light quark mass $m \!=\! 0.56 \beta$, computed numerically via the
BSW-improved Multhopp technique.  All masses are in units of $\beta$.}

\label{t1}

\end{centering}

\end{table}


\begin{table}

\begin{centering}

\begin{tabular}{c|c|c|c|c|c}

$m_Q$ & $M_{H_Q} \!-\! m_Q$ &
$\sqrt{m_Q} \, c_0$ & $m_Q \langle 1\!-\!x \rangle $ &
$m_Q^2 \left( \langle x^2 \rangle \!-\! \langle x \rangle^2 \right)$ &
$\frac{m_Q}{M_{H_Q}} \langle \frac 1 x \rangle \!-\! 1$ \\ \hline

0.26 & 0.81299 & 0.5067 & 0.130 & 0.005 & $3.61 \times
  10^{-1}$ \\

 1.0 & 0.92634 & 0.9481 & 0.373 & 0.050 & $7.60 \times
10^{-2}$ \\

 3.0 & 0.99222 & 1.3788 & 0.658 & 0.202 & $1.40 \times
10^{-2}$ \\

 5.0 & 1.00958 & 1.5420 & 0.772 & 0.306 & $5.40 \times
  10^{-3}$ \\

 7.0 & 1.01725 & 1.6277 & 0.833 & 0.375 & $2.80 \times
  10^{-3}$  \\

10.0 & 1.02288 & 1.6999 & 0.885 & 0.442 & $1.38 \times
  10^{-3}$ \\

15.0 & 1.02692 & 1.7612 & 0.930 & 0.509 & $6.51 \times
  10^{-4}$ \\

25.0 & 1.02946 & 1.8137 & 0.969 & 0.591 & $3.42 \times
  10^{-4}$ \\

35.0 & 1.03003 & 1.8369 & 0.987 & 0.665 & $2.96 \times
  10^{-4}$ \\

50.0 & 1.02992 & 1.8543 & 1.002 & 0.811 & $2.97 \times
  10^{-4}$

\end{tabular}

\caption{The same matrix elements as in Table~\ref{t1} for the ground
state and $m \!=\! 0.26 \beta$.}

\label{t2}

\end{centering}

\end{table}


\begin{table}

\begin{centering}

\begin{tabular}{cc|cc|rc|cc}

$m$ && $M_{H_Q}$ && $\left< \frac{1}{1-x} \right>$ &&
$\frac{m}{M_{H_Q}} \left< \frac{1}{1-x} \right> \!-\! 1$ & \\
\hline \multicolumn{7}{c}{$m_Q \!=\! 7 \beta$} & \\ \hline

0.10 && 7.84710 && 73.0 && $-7.0 \times 10^{-2}$ & \\

0.26 && 8.01725 && 31.1 && $+9.4 \times 10^{-3}$ & \\

0.40 && 8.15422 && 19.2 && $-5.6 \times 10^{-2}$ & \\

0.56 && 8.30423 && 13.8 && $-6.7 \times 10^{-2}$ & \\

1.00 && 8.71728 &&  8.24 && $-5.5 \times 10^{-2}$ & \\

1.50 && 9.19304 &&  5.87 && $-4.3 \times 10^{-2}$ & \\
\hline \multicolumn{7}{c}{$m_Q \!=\! 10 \beta$} & \\ \hline

0.10 && 10.85436 && 98.6 && $-9.2 \times 10^{-2}$ & \\

0.26 && 11.02288 && 42.6 && $+4.5 \times 10^{-3}$ & \\

0.40 && 11.15925 && 26.2 && $-6.2 \times 10^{-2}$ & \\

0.56 && 11.30820 && 18.7 && $-7.4 \times 10^{-2}$ & \\

1.00 && 11.71817 && 11.0 && $-6.1 \times 10^{-2}$ & \\

1.50 && 12.19091 &&  7.73 && $-4.8 \times 10^{-2}$ & \\
\hline \multicolumn{7}{c}{$m_Q \!=\! 20 \beta$} & \\ \hline

0.10 && 20.86224 && 182. && $-1.3 \times 10^{-1}$ & \\

0.26 && 21.02864 &&  81.0 && $+2.0 \times 10^{-3}$ & \\

0.40 && 21.16490 &&  49.4 && $-6.7 \times 10^{-2}$ & \\

0.56 && 21.31284 &&  34.9 && $-8.3 \times 10^{-2}$ & \\

1.00 && 21.71902 &&  20.2 && $-7.0 \times 10^{-2}$ & \\

1.50 && 22.18785 &&  14.0 && $-5.6 \times 10^{-2}$ &

\end{tabular}

\caption{Ground-state matrix elements as functions of $m$, used to
probe the expectation value of the light-quark scalar density
$\matel{H_Q}{\, \bar{q}q \,}{H_Q}/2M_{H_Q}\,$.}

\label{light}

\end{centering}

\end{table}


\begin{table}

\begin{centering}

\begin{tabular}{c|c|c|c|c|c|c}

$k$ & $\epsilon_k$ & $\tau_{k0}$ & $\rho^2 \!
- \! 1/4$ & $\bar \Lambda$ & $\mu_\pi^2$ & $\rho_D^3$ \\ \hline

 0 & 1.318 & --- & --- & --- & --- & --- \\

 1 & 2.516 & $7.25 \times 10^{-1}$ & $6.2 \times 10^{-3}$ & $
1.5 \times 10^{-2}$ & $3.6 \times 10^{-2}$ & $8.8 \times 10^{-2}$
\\

 3 & 3.989 & $5.36 \times 10^{-2}$ & $7.8 \times 10^{-4}$ & $2.7
\times 10^{-3}$ & $9.4 \times 10^{-3}$ & $3.3 \times 10^{-2}$ \\

 5 & 5.060 & $1.73 \times 10^{-2}$ & $2.2 \times 10^{-4}$ & $9.3
\times 10^{-4}$ & $4.1\times 10^{-3}$ & $1.8 \times 10^{-2}$ \\

 7 & 5.949 & $8.37 \times 10^{-3}$ & $8.7 \times 10^{-5}$ & $4.3
\times 10^{-4}$ & $2.2 \times 10^{-3}$ & $1.1 \times 10^{-2}$ \\

 9 & 6.724 & $4.92 \times 10^{-3}$ & $4.1 \times 10^{-5}$ & $2.3
\times 10^{-4}$ & $1.2 \times 10^{-3}$ & $6.7 \times 10^{-3}$ \\

11 & 7.421 & $3.24 \times 10^{-3}$ & $2.1 \times 10^{-5}$ & $1.3
\times 10^{-4}$ & $7.4 \times 10^{-4}$ & $4.3 \times 10^{-3}$ \\

13 & 8.060 & $2.29 \times 10^{-3}$ & $1.1 \times 10^{-5}$ & $7.0
\times 10^{-5}$ & $4.4 \times 10^{-4}$ & $2.7 \times 10^{-3}$ \\

15 & 8.653 & $1.71 \times 10^{-3}$ & $5.4 \times 10^{-6}$ & $3.6
\times 10^{-5}$ & $2.4 \times 10^{-4}$ & $1.5 \times 10^{-3}$ \\

17 & 9.209 & $1.32 \times 10^{-3}$ & $2.1 \times 10^{-6}$ & $1.5
\times 10^{-5}$ & $1.0 \times 10^{-4}$ & $6.6 \times 10^{-4}$ \\

19 & 9.735 & $1.05 \times 10^{-3}$ & --- & --- & --- & ---

\end{tabular}

\caption{Meson mass eigenvalues $\epsilon_k \equiv M_{H_Q}^{(k)} \! -
\! m_Q$ and oscillator strengths $\tau$ as functions of excitation
number $k$, for $m \! = \! 0.56 \beta$.  The value for a given
nonperturbative matrix element for each $k$ indicates the fractional
amount remaining after saturating the corresponding sum rules
Eqs.~(\ref{oscsum1})--(\ref{oscsum4}) by states $n$ with $n \le k$.}

\label{osc}

\end{centering}

\end{table}


\begin{table}

\begin{centering}

\begin{tabular}{cc|cc|cc|cc|c}

$n$ && $M_n$ && $\langle n | H_Q \rangle$ && $\Gamma_n/\Gamma_{H_Q}$
&& $\sum_{m=0}^n \Gamma_m/\Gamma_{H_Q}$ \\ \hline

 0 && 11.3082 && 0.96426 && $9.45 \times 10^{-1}$ && 0.9448313 \\

 1 && 12.5744 && 0.25996 && $5.36 \times 10^{-2}$ && 0.9984744 \\

 2 && 13.4495 && 0.044347 && $1.23 \times 10^{-3}$ && 0.9997062 \\

 3 && 14.1780 && 0.023927 && $2.74 \times 10^{-4}$ && 0.9999801 \\

 4 && 14.8095 && 0.005462 && $1.03 \times 10^{-5}$ && 0.9999903 \\

 5 && 15.3810 && 0.006617 && $9.49 \times 10^{-6}$ && 0.9999998 \\

 6 && 15.9043 && 0.001391 && $1.87 \times 10^{-7}$ && 1.0000000

\end{tabular}

\caption{Illustration of the speed of saturation in excitation number
$n$ of the total hadronic width $\Gamma_{H_Q}$ by partial widths
$\Gamma_n$ from exclusive channels of mass $M_n$.  In this example,
$m_Q \!=\! 15 \beta$, $m_q \!=\! 10 \beta$, and $m \!=\! 0.56 \beta$,
for which $M_{H_Q} \!=\! 16.3113 \beta$.}

\label{satur}

\end{centering}

\end{table}


\begin{table}

\begin{centering}

\begin{tabular}{cc|cc|cc|c}

$m_q$ && $N$ && $\Gamma_{H_Q}/\Gamma_{\rm OPE} \!-\! 1$ &&
$\Gamma_{H_Q}/\Gamma_{Q} \!-\! 1$ \\ \hline

 5.0 && 18 && $1.4 \times 10^{-8}$ && $4.1 \times 10^{-4}$ \\

 6.0 && 16 && $1.2 \times 10^{-7}$ && $4.1 \times 10^{-4}$ \\

 7.0 && 13 && $3.2 \times 10^{-7}$ && $4.1 \times 10^{-4}$ \\

 8.0 && 11 && $6.3 \times 10^{-7}$ && $4.1 \times 10^{-4}$ \\

 9.0 &&  8 && $1.2 \times 10^{-6}$ && $4.2 \times 10^{-4}$ \\

10.0 &&  6 && $2.2 \times 10^{-6}$ && $4.2 \times 10^{-4}$ \\

11.0 &&  4 && $4.3 \times 10^{-6}$ && $4.2 \times 10^{-4}$ \\

12.0 &&  3 && $9.0 \times 10^{-6}$ && $4.2 \times 10^{-4}$ \\

12.5 &&  2 && $2.1 \times 10^{-5}$ && $4.4 \times 10^{-4}$ \\

13.0 &&  1 && $3.2 \times 10^{-5}$ && $4.5 \times 10^{-4}$ \\

13.5 &&  1 && $3.4 \times 10^{-5}$ && $4.5 \times 10^{-4}$ \\

14.0 &&  0 && $5.9 \times 10^{-4}$ && $1.0 \times 10^{-3}$ \\

14.5 &&  0 && $8.4 \times 10^{-4}$ && $1.3 \times 10^{-3}$

\end{tabular}

\caption{Numbers relevant to local duality violation for $m_Q \!=\! 15
\beta$, $m \!=\! 0.56 \beta$, and $m_q$ variable.  $N$ indicates the
excitation number of the heaviest final-state meson kinematically
allowed for the given initial meson mass $M_{H_Q}$: $M_N \!\le \!
M_{H_Q} \!<\! M_{N+1}$.  $\Gamma_{H_Q}$ is the total hadronic width
Eq.~(\ref{140}), while $\Gamma_{\rm OPE}$ and $\Gamma_Q$ are given by
Eqs.~(\ref{141}) and (\ref{Born}), respectively.}

\label{dual15}

\end{centering}

\end{table}


\begin{table}

\begin{centering}

\begin{tabular}{cc|cc|cc|c}

$m_q$ && $N$ && $\Gamma_{H_Q}/\Gamma_{\rm OPE} \!-\! 1$ &&
$\Gamma_{H_Q}/\Gamma_{Q} \!-\! 1$ \\ \hline

5.0 && 6 && $6.9 \times 10^{-6}$ && $9.4 \times 10^{-4}$ \\

5.5 && 5 && $8.6 \times 10^{-6}$ && $9.4 \times 10^{-4}$ \\

6.0 && 4 && $1.3 \times 10^{-5}$ && $9.5 \times 10^{-4}$ \\

6.5 && 3 && $1.8 \times 10^{-5}$ && $9.5 \times 10^{-4}$ \\

7.0 && 3 && $2.1 \times 10^{-5}$ && $9.5 \times 10^{-4}$ \\

7.5 && 2 && $5.5 \times 10^{-5}$ && $9.9 \times 10^{-4}$ \\

8.0 && 1 && $7.6 \times 10^{-5}$ && $1.0 \times 10^{-3}$ \\

8.5 && 1 && $7.9 \times 10^{-5}$ && $1.0 \times 10^{-3}$ \\

9.0 && 0 && $1.4 \times 10^{-3}$ && $2.4 \times 10^{-3}$ \\

9.5 && 0 && $1.9 \times 10^{-3}$ && $2.8 \times 10^{-3}$

\end{tabular}

\caption{Same as in Table~\ref{dual15}, except $m_Q \!=\! 10 \beta$.}

\label{dual10}

\end{centering}

\end{table}


\begin{table}

\begin{centering}

\begin{tabular}{cc|cc|cc|c}

$m_q$ && $N$ && $\Gamma_{H_Q}/\Gamma_{\rm OPE} \!-\! 1$ &&
$\Gamma_{H_Q}/\Gamma_{Q} \!-\! 1$ \\ \hline

1.0 && 3 && $8.5 \times 10^{-5}$ && $3.7 \times 10^{-3}$ \\

1.5 && 3 && $8.0 \times 10^{-5}$ && $3.7 \times 10^{-3}$ \\

2.0 && 2 && $1.9 \times 10^{-4}$ && $3.8 \times 10^{-3}$ \\

2.5 && 2 && $2.5 \times 10^{-4}$ && $3.8 \times 10^{-3}$ \\

3.0 && 1 && $3.5 \times 10^{-4}$ && $3.9 \times 10^{-3}$ \\

3.5 && 1 && $3.2 \times 10^{-4}$ && $3.9 \times 10^{-3}$ \\

4.0 && 0 && $6.6 \times 10^{-3}$ && $1.0 \times 10^{-2}$ \\

4.5 && 0 && $7.3 \times 10^{-3}$ && $1.1 \times 10^{-2}$

\end{tabular}

\caption{Same as in Table~\ref{dual15}, except $m_Q \!=\! 5 \beta$.}

\label{dual5}

\end{centering}

\end{table}


\begin{table}

\begin{centering}

\begin{tabular}{cc|cc|cc|c}

$m_Q$ && $N$ && $\Gamma_{H_Q}/\Gamma_{\rm OPE} \!-\! 1$ &&
$\Gamma_{H_Q}/\Gamma_{Q} \!-\! 1$ \\ \hline

 1.0 &&  0 && $1.3 \times 10^{-1}$ && $1.8 \times 10^{-1}$ \\

 2.0 &&  0 && $2.4 \times 10^{-2}$ && $4.2 \times 10^{-2}$ \\

 3.0 &&  1 && $1.5 \times 10^{-3}$ && $1.1 \times 10^{-2}$ \\

 4.0 &&  2 && $2.7 \times 10^{-4}$ && $5.7 \times 10^{-3}$ \\

 5.0 &&  4 && $6.0 \times 10^{-5}$ && $3.6 \times 10^{-3}$ \\

 6.0 &&  5 && $2.7 \times 10^{-5}$ && $2.6 \times 10^{-3}$ \\

 7.0 &&  7 && $1.1 \times 10^{-5}$ && $1.9 \times 10^{-3}$ \\

 8.0 &&  8 && $5.7 \times 10^{-6}$ && $1.5 \times 10^{-3}$ \\

 9.0 && 10 && $2.8 \times 10^{-6}$ && $1.2 \times 10^{-3}$ \\

10.0 && 13 && $1.3 \times 10^{-6}$ && $9.3 \times 10^{-4}$ \\

11.0 && 15 && $4.5 \times 10^{-7}$ && $7.7 \times 10^{-4}$ \\

12.0 && 18 && $4.8 \times 10^{-8}$ && $6.5 \times 10^{-4}$

\end{tabular}

\caption{Same as in Table~\ref{dual15}, except $m_q \!=\! m \!=\! 0.56 
\beta \,$ fixed and $m_Q$ variable.}

\label{dualmc56}

\end{centering}

\end{table}


\begin{table}

\begin{centering}

\begin{tabular}{cc|cc|cc|c}

$m_Q$ && $N$ && $\Gamma_{H_Q}/\Gamma_{\rm OPE} \!-\! 1$ &&
$\Gamma_{H_Q}/\Gamma_{Q} \!-\! 1$ \\ \hline

 1.0 &&  0 && $2.2 \times 10^{-1}$ && $3.1 \times 10^{-1}$ \\

 2.0 &&  1 && $1.2 \times 10^{-3}$ && $2.9 \times 10^{-2}$ \\

 3.0 &&  1 && $1.1 \times 10^{-3}$ && $1.5 \times 10^{-2}$ \\

 4.0 &&  2 && $1.9 \times 10^{-4}$ && $8.4 \times 10^{-3}$ \\

 5.0 &&  4 && $5.2 \times 10^{-5}$ && $5.5 \times 10^{-3}$ \\

 6.0 &&  5 && $2.4 \times 10^{-5}$ && $3.8 \times 10^{-3}$ \\

 7.0 &&  7 && $1.0 \times 10^{-5}$ && $2.8 \times 10^{-3}$ \\

 8.0 &&  8 && $5.1 \times 10^{-6}$ && $2.2 \times 10^{-3}$ \\

 9.0 && 10 && $2.5 \times 10^{-6}$ && $1.7 \times 10^{-3}$ \\

10.0 && 12 && $1.1 \times 10^{-6}$ && $1.4 \times 10^{-3}$ \\

11.0 && 15 && $4.2 \times 10^{-7}$ && $1.1 \times 10^{-3}$ \\

12.0 && 17 && $5.2 \times 10^{-8}$ && $9.7 \times 10^{-4}$

\end{tabular}

\caption{Same as in Table~\ref{dualmc56}, except $m_q \!=\! m \!=\! 0.26
\beta$.}

\label{dualmc26}

\end{centering}

\end{table}


\begin{table}

\begin{centering}

\begin{tabular}{cc|cc|cc|c}

$n$ && $M_{2n}$ && $\bar R (M_{2n}^2) \!\times\! 10^4$ && $R_0 
(M_{2n}^2) \!\times\! 10^4$ \\ \hline

0 &&  1.7792 && 930.5  && 805.6 \\

1 &&  4.5349 && 15.18  && 15.30 \\

2 &&  6.2796 && 4.082  && 4.099 \\

3 &&  7.6574 && 1.838  && 1.844 \\

4 &&  8.8310 && 1.037  && 1.040 \\

5 &&  9.8703 && 0.6635 && 0.6651 \\

6 && 10.813\hspace{1.0em}  && 0.4602 && 0.4614 \\

7 && 11.681\hspace{1.0em}  && 0.3376 && 0.3385 \\

8 && 12.490\hspace{1.0em}  && 0.2581 && 0.2588 \\

9 && 13.250\hspace{1.0em}  && 0.2036 && 0.2042

\end{tabular}

\caption{Saturation of the vacuum current correlator as depicted
graphically in Fig.~\ref{vacfig}.  The fiducial $q^2$ point in each
interval $M_{2n-1}^2 \!<\! q^2 \!<\! M_{2n+1}^2$, in which $\bar R$ 
(averaged hadronic $\delta$-functions) and $R_0$ (Born-term partonic 
expression) are compared, is chosen to be $M_{2n}^2$.}

\label{vaccor}

\end{centering}

\end{table}


\clearpage

\begin{figure}
  \begin{centering}
  \def\epsfsize#1#2{1.40#2}
  \hfil\epsfbox{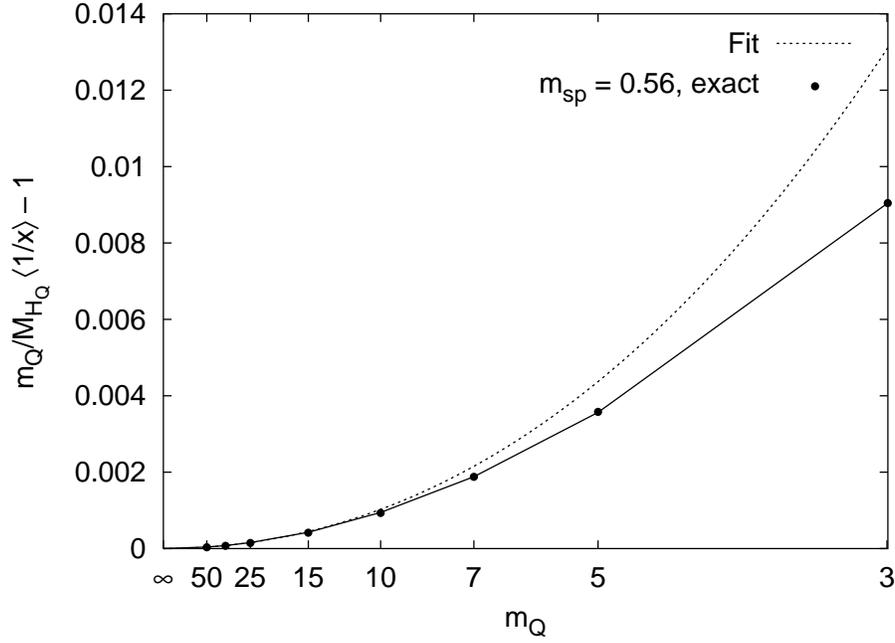}\hfil\hfill

\caption{Values of the ground-state matrix element $(m_Q/M_{H_Q})
\langle 1/x \rangle \!-\! 1$ for $m\!=\!0.56\beta$ (solid line).  The 
dashed line represents an approximation using the the final expansion 
in Eq.~(\ref{58}), good to $O(1/m_Q^3)$.}

\label{Mcond}

\end{centering}
\end{figure}

\end{document}